\pdfoutput=1

\documentclass[11pt,twoside,a4paper,cmspaper,final,collab]{cms-tdr}
\def\svnVersion{075ca32-D}\def\svnDate{2019/11/14}\def\cmsCernNoTag{CERN-EP-2018-321}\def\cmsCernDate{\today}\def\cmsMessage{"Published in Physics Letters B as \href{http://dx.doi.org/10.1016/j.physletb.2019.135042}{\doi{10.1016/j.physletb.2019.135042}.}"}

\begin{document}\cmsNoteHeader{TOP-17-011}

\hyphenation{had-ron-i-za-tion}
\hyphenation{cal-or-i-me-ter}
\hyphenation{de-vices}
\RCS$HeadURL$
\RCS$Id$

\cmsNoteHeader{TOP-17-011}
\title{Measurement of the single top quark and antiquark production cross sections in the $t$ channel and their ratio in proton-proton collisions at $\sqrt{s}=13\TeV$}

\newcommand{\sigmatchtop}{\ensuremath{\sigma_{t\text{-ch,}\cPqt}}\xspace}
\newcommand{\sigmatchantitop}{\ensuremath{\sigma_{t\text{-ch,}\cPaqt}}\xspace}
\newcommand{\sigmatchtotal}{\ensuremath{\sigma_{t\text{-ch,}\cPqt+\cPaqt}}\xspace}

\newcommand{\sigmatchtotalvtb}{\ensuremath{\sigma^{}_{t\text{-ch,}\cPqt+\cPaqt}}\xspace}
\newcommand{\sigmatchtotalvtbtheo}{\ensuremath{\sigma^{\text{theo}}_{t\text{-ch,}\cPqt+\cPaqt}}\xspace}

\newcommand{\Rt}{\ensuremath{R_{t\text{-ch}}}\xspace}
\newcommand{\vtb}{\ensuremath{V_{\cPqt\cPqb}}\xspace}
\newcommand{\flv}{\ensuremath{f_{\mathrm{LV}}}\xspace}

\newcommand{\murmuf}{\ensuremath{\mu_\mathrm{R}/\mu_\mathrm{F}}\xspace}
\newcommand{\tW}{\ensuremath{\cPqt\PW}\xspace}

\newcommand{\mtop}{\ensuremath{m_{\cPqt}}\xspace}
\newcommand{\costhetastar}{\ensuremath{\cos\theta^{*}}\xspace}

\newcommand{\PFrelIso}{\ensuremath{I_{\text{rel}}}\xspace}
\newcommand{\mtw}{\ensuremath{\mT^\PW}\xspace}
\newcommand{\rtwojonet}{2jets-1tag\xspace}
\newcommand{\rtwojzerot}{2jets-0tags\xspace}
\newcommand{\rthreejonet}{3jets-1tag\xspace}
\newcommand{\rthreejtwot}{3jets-2tags\xspace}

\newcommand{\xstopresult}{130\xspace}
\newcommand{\xstopstat}{1\xspace}
\newcommand{\xstopprofiledunc}{4\xspace}
\newcommand{\xstoplumiunc}{3\xspace}
\newcommand{\xstopextunc}{18\xspace}
\newcommand{\xstopsysunc}{19\xspace}
\newcommand{\xstoptotalunc}{19\xspace}

\newcommand{\xsantitopresult}{77\xspace}
\newcommand{\xsantitopstat}{1\xspace}
\newcommand{\xsantitopprofiledunc}{2\xspace}
\newcommand{\xsantitoplumiunc}{2\xspace}
\newcommand{\xsantitopextunc}{11\xspace}
\newcommand{\xsantitopsysunc}{12\xspace}
\newcommand{\xsantitoptotalunc}{12\xspace}

\newcommand{\rtopantitopresult}{1.68\xspace}
\newcommand{\rtopantitopstat}{0.02\xspace}
\newcommand{\rtopantiprofiledunc}{0.02\xspace}
\newcommand{\rtopantiextunc}{0.05\xspace}
\newcommand{\rtopantisysunc}{0.05\xspace}
\newcommand{\rtopantitotalunc}{0.06\xspace}

\newcommand{\xstotalresult}{207\xspace}
\newcommand{\xstotalstat}{2\xspace}
\newcommand{\xstotalprofiledunc}{6\xspace}
\newcommand{\xstotallumunc}{5\xspace}
\newcommand{\xstotalextunc}{29\xspace}
\newcommand{\xstotalsysunc}{31\xspace}
\newcommand{\xstotaltotalunc}{31\xspace}

\newcommand{\vtbresult}{0.98\xspace}
\newcommand{\vtbexpunc}{0.07\xspace}

\date{\today}

\abstract{Measurements of the cross sections for the production of single top quarks and antiquarks in the $t$ channel, and their ratio, are presented for proton-proton collisions at a center-of-mass energy of 13\TeV. The data set used was recorded in 2016 by the CMS detector at the LHC and corresponds to an integrated luminosity of 35.9\fbinv. Events with one muon or electron are selected, and different categories of jet and \cPqb\ jet multiplicity and multivariate discriminators are applied to separate the signal from the background. The cross sections for the $t$-channel production of single top quarks and antiquarks are measured to be $\xstopresult\pm\xstopstat\stat\pm\xstopsysunc\syst\unit{pb}$ and $\xsantitopresult\pm\xsantitopstat\stat\pm\xsantitopsysunc\syst\unit{pb}$, respectively, and their ratio is $\rtopantitopresult\pm\rtopantitopstat\stat\pm\rtopantisysunc\syst$. The results are in agreement with the predictions from the standard model.}

\hypersetup{%
pdfauthor={CMS Collaboration},%
pdftitle={Measurement of the single top quark and antiquark production cross sections in the t channel and their ratio in proton-proton collisions at sqrt(s)=13 TeV},%
pdfsubject={CMS},%
pdfkeywords={CMS, physics, top quark, single top, cross section}}

\maketitle

\section{Introduction}
\label{sec:introduction}

\begin{figure*}[h!]
\centering
\includegraphics[width=0.3\textwidth]{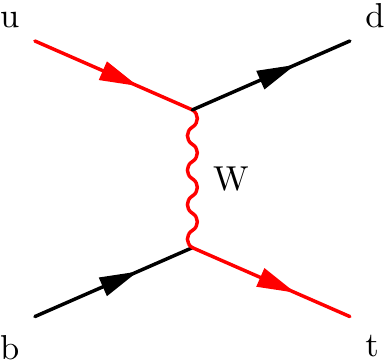} \hspace{3em}
\includegraphics[width=0.3\textwidth]{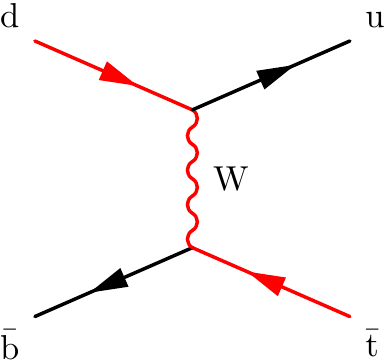}
\caption{\label{fig:Feynmangraph} Feynman diagrams at Born level for the electroweak production of a single top quark (left) and antiquark (right). The flavor of the light quark in the initial state---either up quark (\cPqu) or down quark (\cPqd)---defines whether a top quark or top antiquark is produced.}
\end{figure*}

The study of single top quark production provides important insight into the electroweak processes of the standard model (SM) of elementary particles and into the structure of the proton.
It also provides access to the magnitude of the Cabibbo--Kobayashi--Maskawa (CKM) matrix element \vtb.
Among the production channels, the $t$-channel process is the dominant mechanism in proton-proton (\Pp{}\Pp) collisions at the CERN LHC accounting for approximately 70$\%$ of the total single top quark production cross section at $\sqrt{s} = 13\TeV$~\cite{HUSEMANN201748}.
The $t$ channel has a very distinct signature with a light quark, which is predominantly produced in the forward direction, and a top quark.
Figure~\ref{fig:Feynmangraph} illustrates the production of a single top quark and a single top antiquark.
The flavor of the initial light quark defines the charge of the produced top quark; up quarks in the initial state result in top quarks, while down quarks produce top antiquarks.
The ratio of the cross sections of these two processes provides insight into the inner structure of the proton as described by the parton distribution functions (PDFs).
The ATLAS and CMS Collaborations have performed several measurements of the cross section for single top quark production in the $t$ channel using LHC data collected at $\sqrt{s} = 7$, 8, and 13\TeV \cite{Aad:2012ux, Aad:2014fwa, Aaboud:2017pdi, Aaboud:2016ymp, Chatrchyan:2011vp, Chatrchyan:2012ep, Khachatryan:2014iya, Sirunyan:2016cdg}.
With a data set corresponding to an integrated luminosity of 35.9\fbinv, the analysis described in this letter uses about 18 times more data compared to the previous analysis at 13\TeV~\cite{Sirunyan:2016cdg} and also exploits the electron final state.

The 13\TeV $t$-channel single top quark cross section has been calculated to next-to-leading-order (NLO) accuracy in quantum chromodynamics (QCD) using $\textsc{hathor}$ 2.1 \cite{Aliev:2010zk, Kant:2014oha}. Assuming a top quark mass of 172.5\GeV and $\vtb=1$, the calculation yields cross section values of
\begin{linenomath}
\begin{equation}
\begin{aligned} \label{eq:prediction}
\sigmatchtop &=& 136.0^{+4.1}_{-2.9} (\text{scale}) \pm 3.5 (\mathrm{PDF +} \alpS) \unit{pb}, \\
\sigmatchantitop &=& 81.0^{+2.5}_{-1.7} (\text{scale}) \pm 3.2 (\mathrm{PDF +} \alpS) \unit{pb}, \\
\sigmatchtotal &=& 217.0^{+6.6}_{-4.6} (\text{scale}) \pm 6.2 (\mathrm{PDF +} \alpS) \unit{pb},
\end{aligned}
\end{equation}
\end{linenomath}
for the $t$-channel production of single top quarks ($\sigmatchtop$), single top antiquarks ($\sigmatchantitop$), and the sum of both subprocesses ($\sigmatchtotal$), respectively, where \alpS is the strong coupling constant. The cross sections are evaluated in the five-flavor scheme (5FS), where the \cPqb\ quark is described by the PDF of the incoming protons. The quoted uncertainties are associated with the renormalization and factorization scales, as well as \alpS at the mass of the \PZ boson, and PDFs. The PDF and $\alpS(m_\PZ)$ uncertainties were calculated with the MSTW2008 68\% confidence level NLO \cite{Martin:2009iq,Martin:2009bu}, CT10 NLO \cite{Lai:2010vv}, and NNPDF2.3 \cite{Ball:2012cx} PDF sets, using the PDF4LHC prescription \cite{Botje:2011sn, Alekhin:2011sk}. A prediction at full next-to-next-to-leading-order (NNLO) accuracy~\cite{Berger:2016oht} is also available for single top quark production in the $t$ channel at 13\TeV. As the available NNLO calculations consider only uncertainties from variations in the renormalization and factorization scales, the NLO prediction providing all required systematic uncertainty sources is used instead in this analysis for the normalization of the signal process.
Depending on the PDF set used, the predicted values for the cross sections for the two processes and their ratio may differ, rendering the measurement sensitive to various PDF parameterizations.
Using the cross section values and the PDF sets given above, the predicted value for the ratio $\Rt=\sigmatchtop/\sigmatchantitop$ is $1.68 \pm 0.08$, where the uncertainty includes contributions due to variations of the renormalization and the factorization scales, the top quark mass, and the PDF and \alpS.

The analysis uses events containing a single isolated muon or electron in the final state. The muon or electron originates from the decay of the \PW boson from the top quark decay, either directly or via $\PW\to\tau\nu$ and the following $\tau\to\ell\nu$ decay, where $\ell$ refers to either a muon or an electron. The main backgrounds come from the production of top quark-antiquark pairs (\ttbar) and from the production of \PW bosons in association with jets ($\PW$+jets). The separation between signal and background is achieved using boosted decision trees (BDTs), which combine the discriminating power of several kinematic distributions into a single classifier. The cross sections of $t$-channel single top quark and single top antiquark production, as well as the ratio of the two processes, are determined from a fit to the distributions of this single classifier.

\section{The CMS detector}
\label{sec:detector}
The central feature of the CMS apparatus is a superconducting solenoid of 6\unit{m} internal diameter, providing a
magnetic field of 3.8\unit{T}. Within the solenoid volume are a silicon pixel and strip tracker, a lead tungstate
crystal electromagnetic calorimeter (ECAL), and a brass and scintillator hadron calorimeter, each composed
of a barrel and two endcap sections. Forward calorimeters extend the pseudorapidity ($\eta$) coverage provided by the barrel
and endcap detectors. Muons are measured in the range $\abs{\eta} < 2.4$, with detection planes made using three technologies: drift tubes, cathode strip chambers, and resistive plate chambers, embedded in the steel flux-return yoke outside the solenoid. The electron momentum is estimated by combining the energy measurement in the ECAL with the momentum measurement in the tracker. Events of interest are selected using a two-tiered trigger system~\cite{Khachatryan:2016bia}. The first level, composed of custom hardware processors, uses information from the calorimeters and muon detectors to select events. The second level, known as the high-level trigger, consists of a farm of processors running a version of the full event reconstruction software optimized for fast processing. A more detailed description of the CMS detector, together with a definition of the coordinate system used and the relevant kinematic variables, can be found in Ref.~\cite{Chatrchyan:2008zzk}.

\section{Simulation of events}
\label{sec:simulation}

Signal and background events are simulated to NLO accuracy with either the \POWHEG or the \MGvATNLO~\cite{Alwall:2014hca} Monte Carlo (MC) event generator. The $t$-channel signal process~\cite{Alioli:2009je} is simulated with \POWHEG 2.0 \cite{Nason:2004rx, Frixione:2007vw,Alioli:2010xd} in the four-flavor scheme (4FS),  where \cPqb\ quarks are produced via gluon splitting. This scheme yields a more precise description of the kinematic distributions of $t$-channel signal events than the 5FS~\cite{Frederix:2012dh,Aaboud:2017pdi}. For the normalization of the signal samples the predictions derived in the 5FS (see Eq.~(\ref{eq:prediction})) are employed. The \ttbar background~\cite{Alioli:2011as} is simulated using \POWHEG 2.0 and is normalized to the prediction calculated with \textsc{top++} 2.0~\cite{Czakon:2011xx}. The production of single top quarks associated to \PW bosons (\tW) is simulated with \POWHEG 1.0 in the 5FS~\cite{Re:2010bp}, normalized to a prediction providing approximate NNLO accuracy~\cite{Kidonakis:2010ux,Kidonakis:2013zqa}. The value of the top quark mass in the simulated samples is $\mtop = 172.5\GeV$.  Events with \PW and \PZ bosons in association with jets are simulated using \MGvATNLO 2.2.2 and the FxFx merging scheme \cite{Frederix:2012ps}. Predictions calculated with \FEWZ 3.1~\cite{Gavin:2010az,Gavin:2012sy,Li:2012wna} are employed for the normalization of these two processes. For all samples, \PYTHIA 8.212~\cite{Sjostrand:2014zea} is used to simulate parton shower and hadronization.
The underlying event is modeled for all samples using the tune CUETP8M1~\cite{Khachatryan:2015pea}, except for the \ttbar sample, for which the tune CUETP8M2T4~\cite{CMS:2016kle} is used, which provides a more accurate description of the kinematic distributions of the top quark pair and of the jets in \ttbar events. The parameterization of the PDFs used in all simulations is NNPDF3.0 NLO~\cite{Ball:2014uwa}. All of the generated events undergo a full simulation of the detector response using a model of the CMS detector implemented in \GEANTfour~\cite{Agostinelli:2002hh}. Additional \Pp{}\Pp interactions within the same or nearby bunch crossing (pileup) are included in the simulation with the same distribution as observed in data.

\section{Event selection and top quark reconstruction}
\label{sec:selection}

In this analysis, the signature of the single top quark $t$-channel production process consists of a charged lepton, a neutrino, which is observed as \pt imbalance, a light-quark jet, which is often produced in the forward direction, and a jet arising from the hadronization of a bottom quark (\cPqb\ jet) from the top quark decay. A second \cPqb\ jet, arising in the production process via gluon splitting, generally has a softer \pt spectrum and a broader $\eta$ distribution compared to the \cPqb\ jet originating from the top quark decay, and therefore often escapes detection. The event selection criteria are chosen according to this signature and events must contain one muon or electron candidate and at least two jets. Events in the muon channel are selected online using a trigger that requires an isolated muon with $\pt > 24\GeV$ and $\abs{\eta} < 2.4$. In the electron channel, a trigger is used that requires electrons with $\pt > 32\GeV$ and $\abs{\eta} < 2.1$.
Only events for which at least one primary vertex is reconstructed are considered in the analysis. The primary vertex must be reconstructed from at least four tracks that have a longitudinal distance $\abs{d_{z}}<24\cm$ and a radial distance $\abs{d_{xy}}<2\cm$ from the interaction point. If more than one primary vertex is found in an event, the reconstructed vertex with the largest value of summed physics-object $\pt^2$ is taken to be the primary \Pp{}\Pp interaction vertex. The physics objects are the jets, clustered using the jet finding algorithm~\cite{Cacciari:2008gp,Cacciari:2011ma} with the tracks assigned to the vertex as inputs, and the associated missing transverse momentum, taken as the negative vector sum of the \pt of those jets.

The particle-flow (PF) algorithm~\cite{CMS-PRF-14-001}, which optimally combines information from all subdetectors, is used for the reconstruction of the individual particles. Muon candidates must have at least one hit in the muon detector and at least five hits in the silicon tracker. They are then reconstructed by a global fit to the information from the silicon tracker and the muon spectrometer. Selected muons must fulfill the criteria $\pt>26\GeV$, $\abs{\eta}<2.4$, and relative isolation, $\PFrelIso < 0.06$. The $\PFrelIso$ of a charged lepton candidate is calculated by summing the transverse energy deposited by photons and charged and neutral hadrons within a cone of size $\sqrt{\smash[b]{(\Delta\eta)^2+(\Delta\phi)^2}} < 0.4$ (where $\phi$ is the azimuthal angle in radians) for muons and 0.3 for electrons around its direction, corrected for contributions from pileup~\cite{Sirunyan:2018}, relative to its \pt.

Electron candidates are reconstructed by fitting tracks in the silicon tracker using the Gaussian-sum filter~\cite{reco_el} and matching the tracks to energy clusters in the ECAL. The electron identification is performed using nine different variables and various selection criteria, including a requirement on the relative isolation $\PFrelIso < 0.06$. Electrons are required to have $\pt > 35\GeV$ and $\abs{\eta} < 2.1$, while electrons falling in the gap between the ECAL barrel and endcap regions ($1.44 < \abs{\eta} < 1.57$) are rejected.
Events containing additional muons with $\pt > 10\GeV$ and $\abs{\eta} < 2.4$ or additional electrons with $\pt > 15\GeV$ and $\abs{\eta} < 2.5$  are rejected. In both cases, the criteria on the lepton identification and isolation are relaxed ($\PFrelIso < 0.2$ for muons; $\PFrelIso < 0.18$ for electrons in the ECAL barrel and $\PFrelIso < 0.16$ for electrons in the ECAL endcaps). Lepton \pt- and $\eta$-dependent scale factors are applied to correct for differences in the lepton reconstruction efficiencies between data and simulation.

Jets are clustered using the anti-\kt clustering algorithm~\cite{Cacciari:2008gp} with a distance parameter of 0.4, as implemented in the \FASTJET package~\cite{Cacciari:2011ma}. The effect of additional tracks and calorimetric energy deposits from pileup on the jet momentum is mitigated by discarding tracks identified to be originating from pileup vertices and applying an offset correction to account for remaining contributions. Jet energy corrections are derived from simulation to bring the measured average response of jets to that of particle-level jets. In situ measurements of the momentum balance in dijet, $\PGg$+jet, $\PZ$+jet, and multijet events are used to account for any residual differences in the jet energy scale in data and simulation~\cite{Khachatryan:2016kdb}. In this analysis, jets with $\pt > 40\GeV$ and $\abs{\eta} < 4.7$ are selected. The combined secondary vertex algorithm (CSVv2)~\cite{Sirunyan:2017ezt} is used to identify \cPqb\ jets, which are required to have $\pt > 40\GeV$ and $\abs{\eta} < 2.4$. The efficiency to identify jets from \cPqb\ quarks is about 40\% at the chosen working point, while the probability to misidentify jets from light quarks or gluons as \cPqb\ jets is 0.1\%. Corrections to the simulation are applied in order to account for the difference in the \cPqb\ tagging efficiency in data and simulation.

To suppress the background from QCD multijet processes in the electron channel, events must fulfill $\ptmiss > 30\GeV$. For events in the muon channel this variable is not sufficiently well modelled, and a requirement on the transverse mass of the \PW boson is imposed instead. The transverse mass of the \PW boson is defined as
\begin{linenomath}
    \begin{equation}
      \mtw = \sqrt{\left(2p_{\mathrm{T},\mu}\ptmiss\right)\left(1-\cos\Delta\phi\right)}  > 50\GeV.
    \end{equation}
\end{linenomath}
Here, \ptmiss is the magnitude of the transverse momentum vector \ptvecmiss. This vector is defined as the projection onto the plane perpendicular to the beam axis of the negative vector sum of the momenta of all reconstructed PF objects in an event. The energy scale corrections applied to jets are propagated to \ptvecmiss~\cite{CMS-PAS-JME-16-004}. The angle between the directions of the momentum vector of the muon and \ptvecmiss is $\Delta\phi$.

The selected events are divided into four different categories, depending on the number of selected jets and the number of \cPqb-tagged jets ($n$jets-$m$tags).
The category with two selected jets, one of which is identified as originating from a bottom quark, \ie, \rtwojonet category, provides the largest contribution of signal events constituting the signal category.
Events with three selected jets with one or two of them \cPqb-tagged, namely events in the \rthreejonet and \rthreejtwot categories, are dominated by \ttbar production.
These serve as control categories that are used in the fit to constrain the contribution from this dominant background process.
Besides these categories, a fourth category containing events with two selected jets and no identified \cPqb\ jets, the \rtwojzerot category, is defined to validate the estimation of the QCD multijet background contribution in data.

\begin{figure*}[h!]
\centering
    \includegraphics[width=0.49\textwidth]{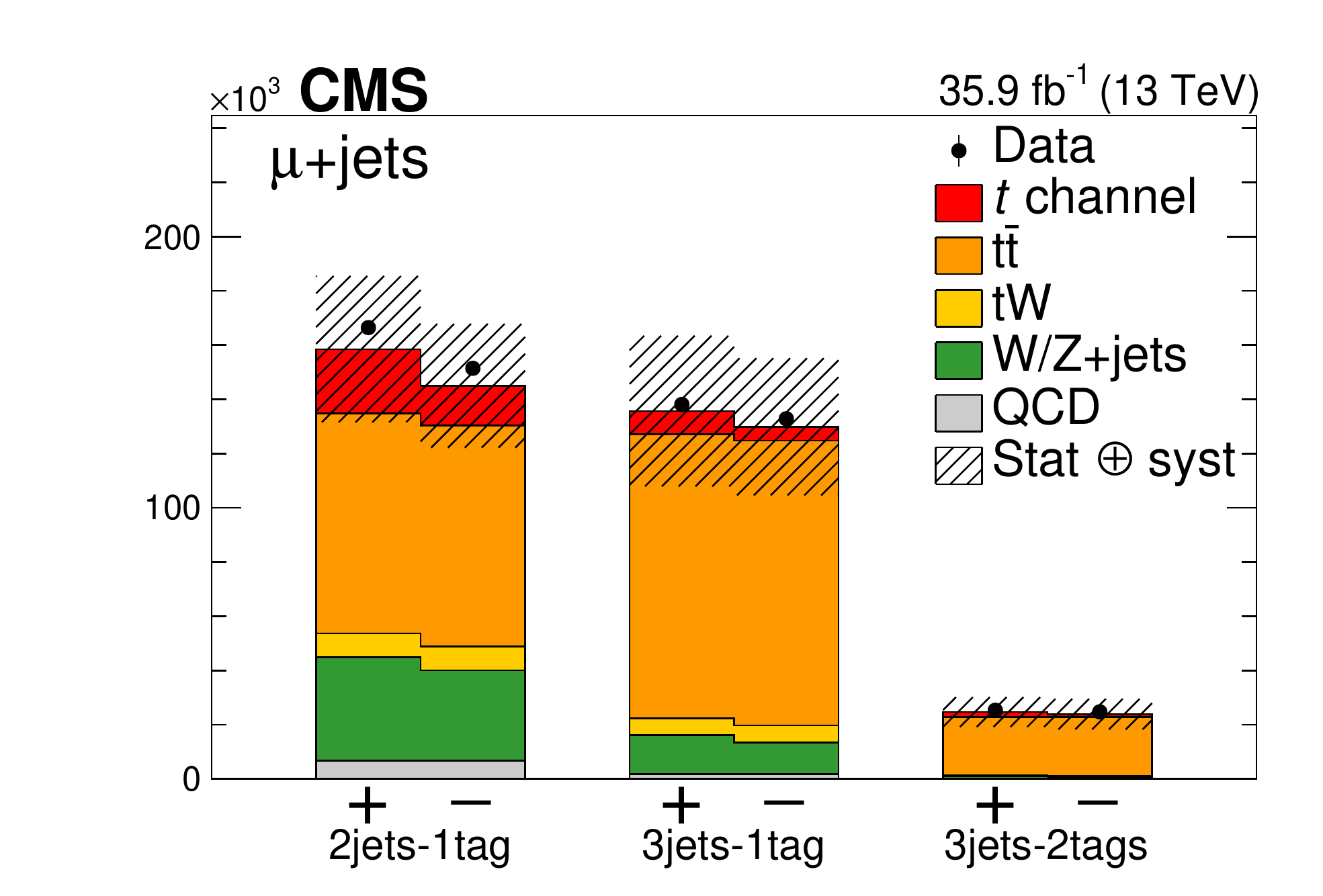}
    \includegraphics[width=0.49\textwidth]{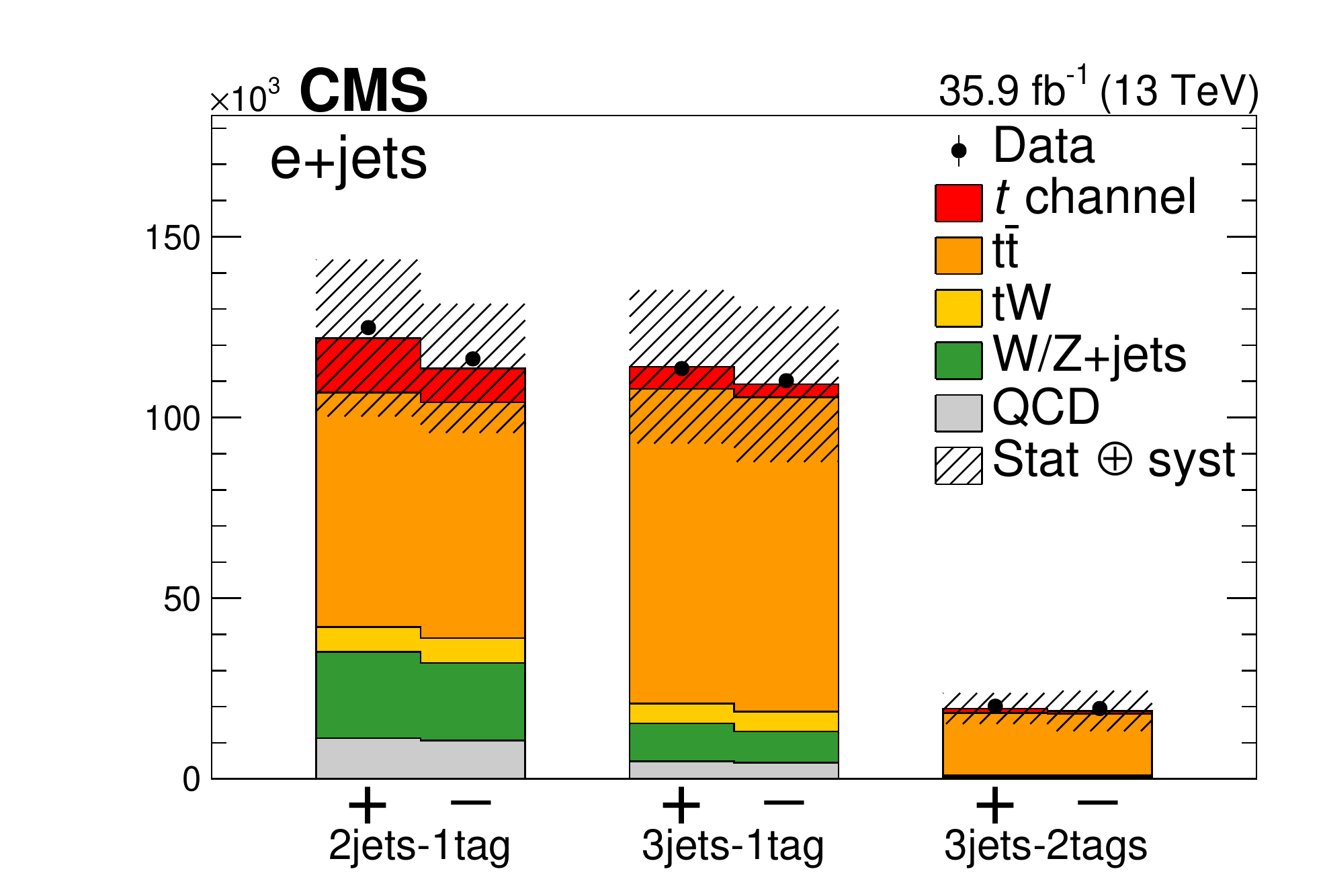}
    \caption{Event yields for the relevant processes in all categories after applying the full event selection in the muon (left) and electron (right) channels.
            The yields are shown separately for positively ($+$) and negatively ($-$) charged muons (electrons).
            The uncertainties include statistical and all systematic uncertainties.
            The yields are obtained from simulation, except for the QCD multijet contribution, which is derived from data (see Section~\ref{sec:bkgmodel}).
}
\label{fig:eventYield}
\end{figure*}

The numbers of selected events are shown in Fig.~\ref{fig:eventYield} for the muon and electron channels.
In both channels, the event yields are shown separately for events with positively and negatively charged muons (electrons).
Positively charged leptons stem from top quarks and negatively charged leptons from top antiquarks.
The contribution from the QCD multijet background is determined directly from data as described in Section~\ref{sec:bkgmodel}. For the other processes, the event yields are derived from simulation.

The momentum four-vector of the top quark is reconstructed from the momenta of its decay products: the charged lepton, the reconstructed neutrino, and the \cPqb\ jet. The ambiguity of the assignment of one of the two \cPqb-tagged jets to the \cPqb\ quark from the top quark decay in the \rthreejtwot category is solved by choosing the \cPqb\ jet that leads to a reconstructed top quark mass closer to the top quark mass in the simulation. In the \rthreejonet category, two untagged jets exist, of which the one with the highest $\abs{\eta}$ is assigned to the light-quark jet in forward direction. The transverse momentum of the neutrino, $\vec{p}_{\mathrm{T},\nu}$, can be obtained from \ptvecmiss. Assuming energy-momentum conservation at the $\PW\ell\nu$ vertex and setting the \PW boson mass to $m_{\PW} = 80.4\GeV$, the longitudinal momentum of the neutrino, $p_{z,\nu}$, can be calculated as
\begin{linenomath}
    \begin{equation}
    \label{eq:nusolver}
        p_{z,\nu} =\frac{\Lambda p_{z,\ell}}{p_{\mathrm{T},\ell}^2}\pm\frac{1}{p_{\mathrm{ T},\ell}^2}\sqrt{\Lambda^2 p_{z,\ell}^2-p_{\mathrm{T},\ell}^2(p_{\ell}^{2} p_{\mathrm{T},\nu}^2-\Lambda^2)},
    \end{equation}
\end{linenomath}
where
\begin{linenomath}
    \begin{equation}
    \label{eq:lambda}
        \Lambda=\frac{m_{\mathrm{W}}^2}{2}+\vec{p}_{\mathrm{T},\ell}\cdot\vec{p}_{\mathrm{T},\nu},
    \end{equation}
\end{linenomath}
and $p_{\ell}^{2}=p_{\mathrm{T},\ell}^2 + p_{z,\ell}^2$ denotes the squared momentum of the charged lepton. In general, this procedure results in two possible solutions for $p_{z,\nu}$, which can have either real or complex values. If both solutions take real values, the one with the smallest absolute value is chosen~\cite{Abazov:2009ii,Aaltonen:2009jj}. In the case of complex solutions, the transverse components of the neutrino momentum are modified such that the algebraic discriminant in Eq.~(\ref{eq:nusolver}) becomes null, while still fulfilling the constraint on the \PW boson mass. Of the possible solutions for $p_{x,\nu}$ and $p_{y,\nu}$ that resolve the problem of the negative discriminant, the coordinate pair that is closest to the corresponding components of \ptvecmiss is chosen.

\section{Modeling and normalization of the QCD multijet background}
\label{sec:bkgmodel}

\begin{figure*}
\centering
\includegraphics[width=0.45\textwidth]{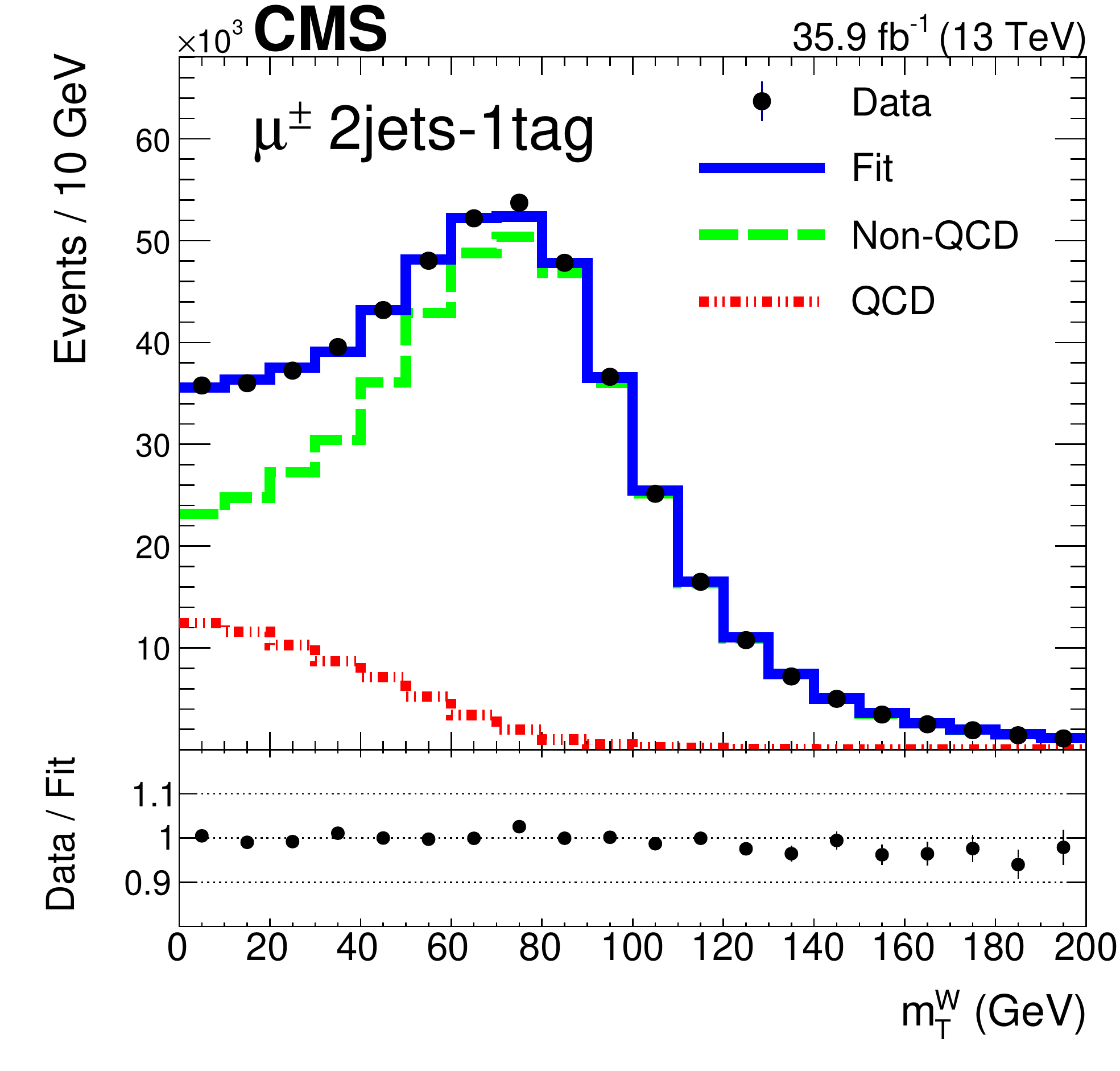}
\includegraphics[width=0.45\textwidth]{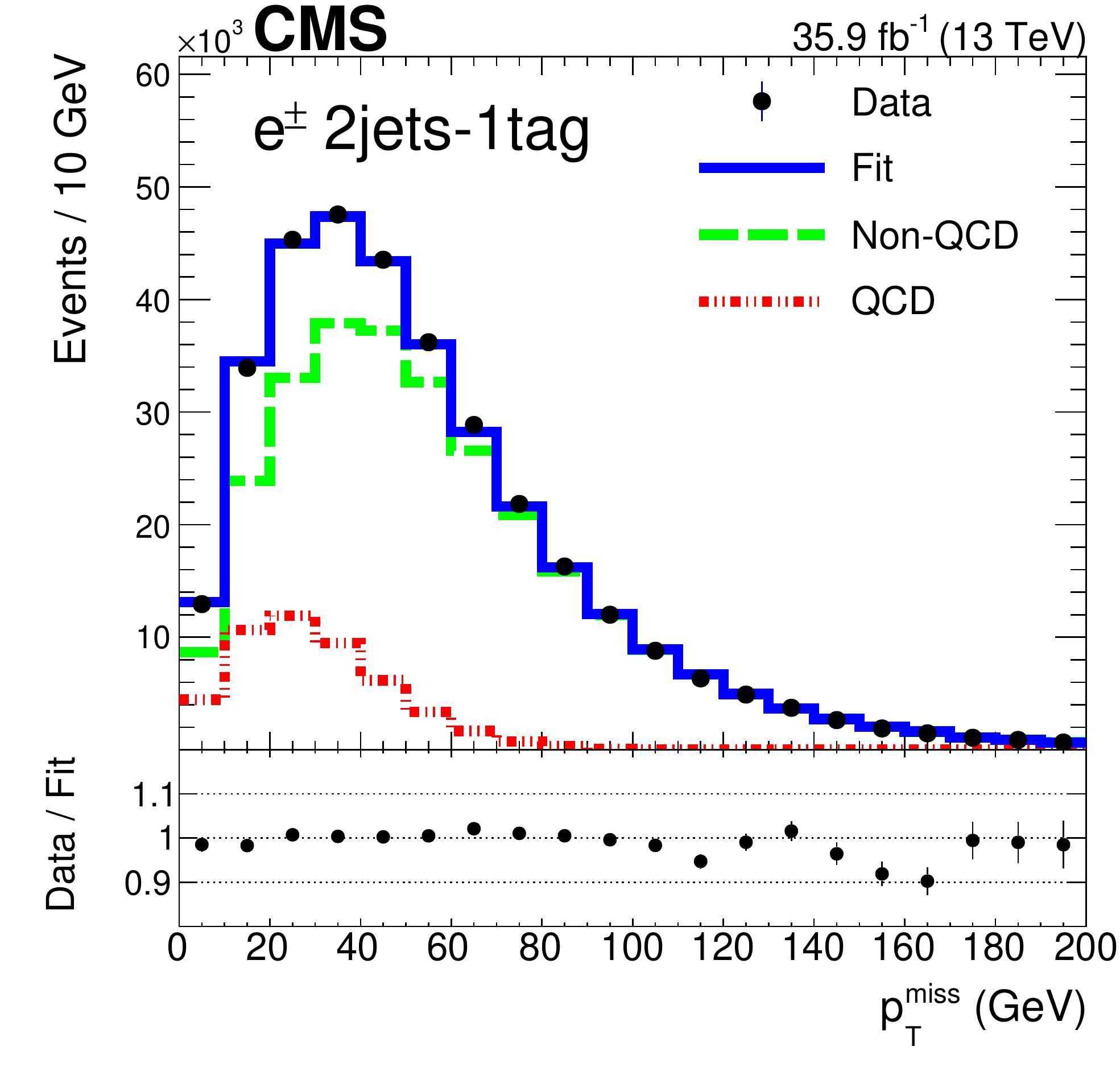}\\
\includegraphics[width=0.45\textwidth]{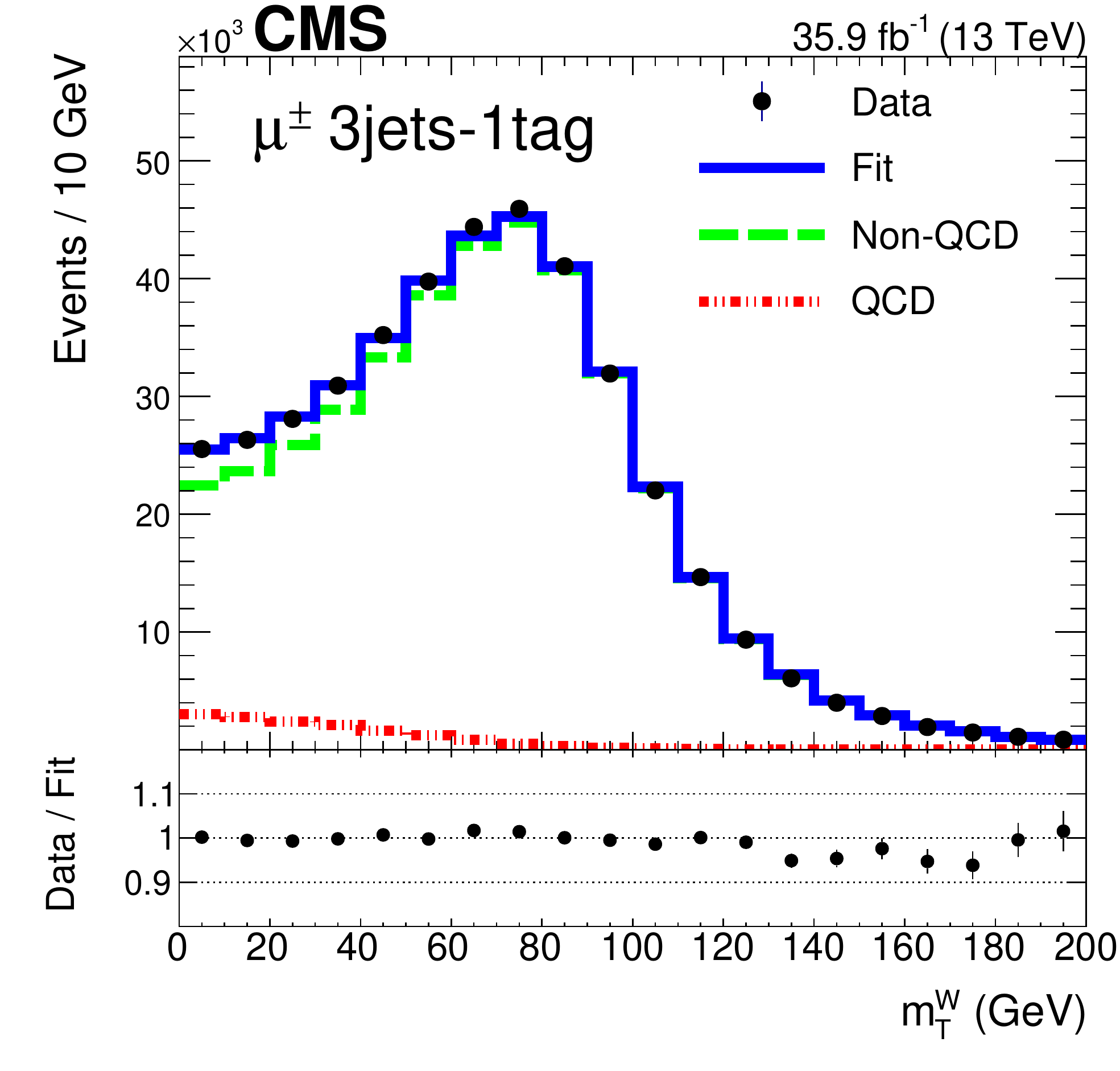}
\includegraphics[width=0.45\textwidth]{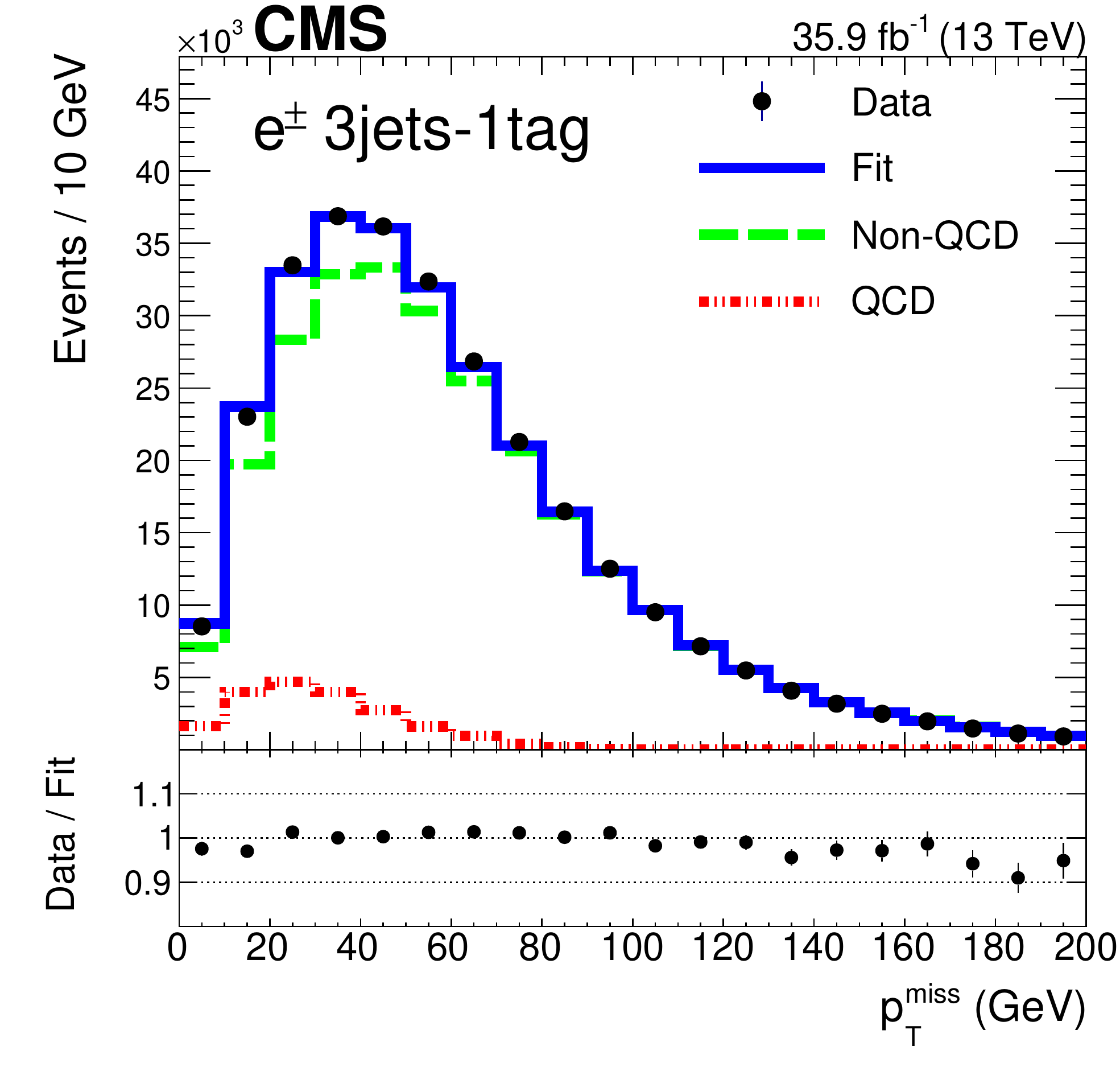} \\
\includegraphics[width=0.45\textwidth]{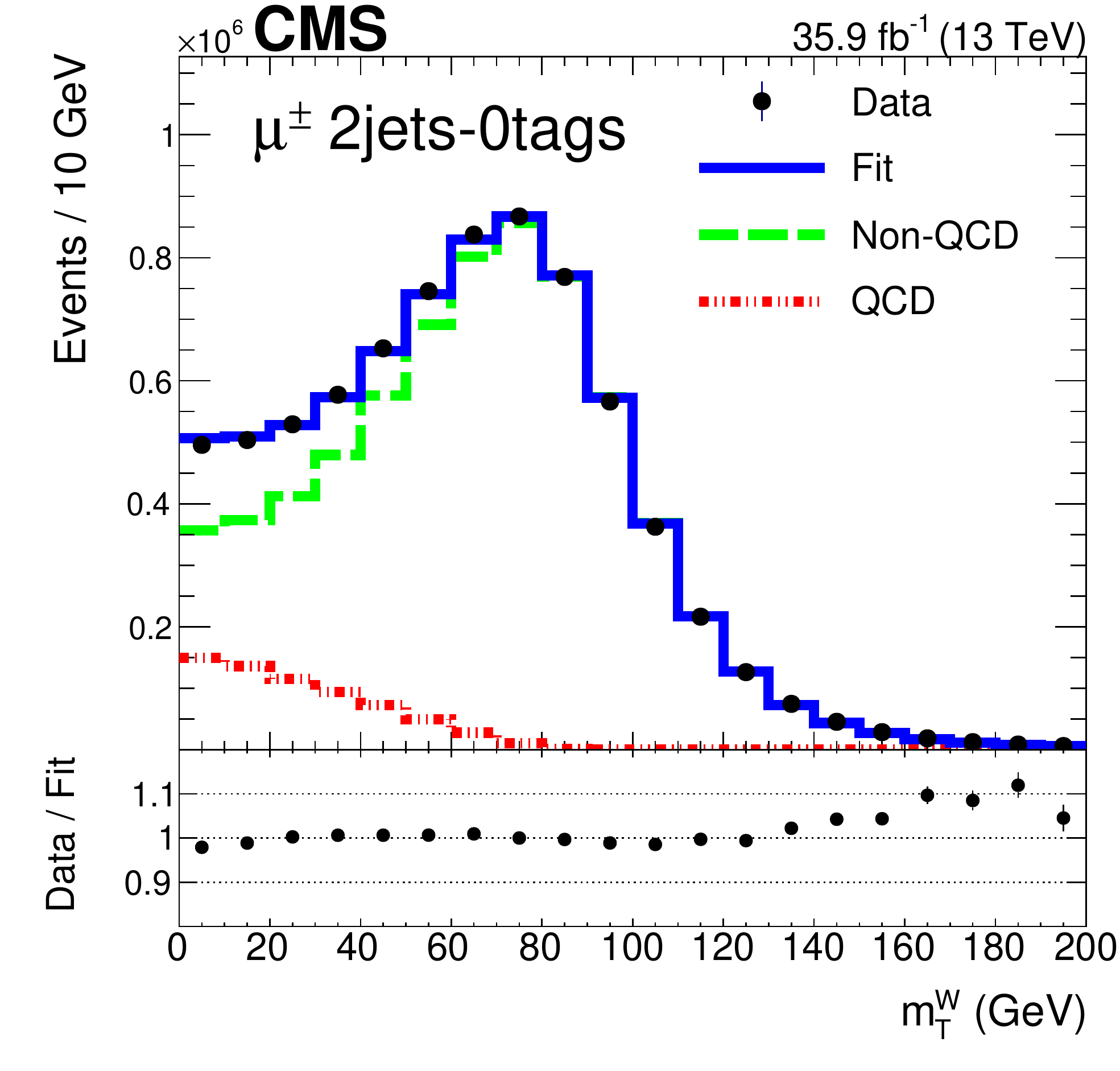}
\includegraphics[width=0.45\textwidth]{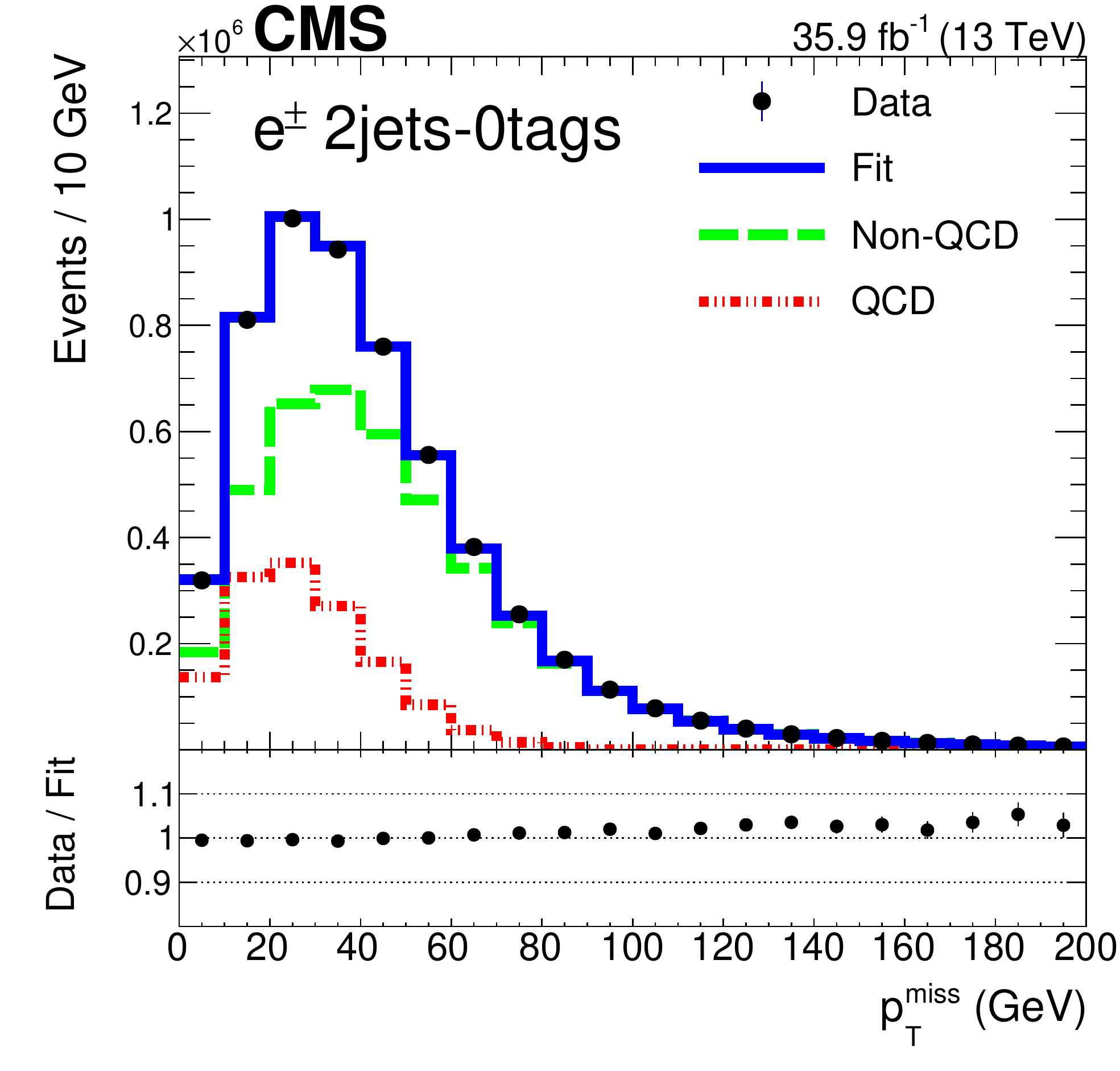} \\

\caption{\label{fig:QCDestimation} Outcome of the maximum likelihood fit to the \mtw distribution of events with  muons (left) and to the \ptmiss distribution for events with electrons (right) in the \rtwojonet (upper row), \rthreejonet (middle row), and the \rtwojzerot (lower row) categories.
The QCD background template is extracted from the sideband region in data. For the fit, only statistical uncertainties are considered.}
\end{figure*}

Because of the theoretically challenging simulation of QCD multijet processes, this background contribution is suppressed as much as possible and the remaining contamination is modeled from data. As described in Section~\ref{sec:selection}, requirements on \mtw or \ptmiss are applied on the events in the muon and electron channels to suppress events from QCD multijet production. Different variables have been chosen for the two channels as \mtw is found to be better modeled compared to \ptmiss in the muon channel and vice versa in the electron channel. These variables provide the highest separation power between QCD multijet events and other processes, including the $t$-channel single top quark production, for the respective lepton final state. The remaining QCD contribution is modeled with samples of events derived from sideband regions in data enriched in QCD multijet events. These sideband regions are defined by inverting the muon or electron isolation requirements, while all other selection criteria described in Section~\ref{sec:selection} remain the same. As no reliable prediction for the QCD contribution to the selected data is available, the normalization for the QCD modeling samples are estimated from data. For that purpose, the same variables that are used to suppress this background contributions are explored by fitting their distributions over the entire range. A maximum likelihood fit is performed on the \mtw or \ptmiss distribution using two probability density functions, one for the QCD multijet process and one for all other non-QCD processes. The latter distribution is obtained by adding the different non-QCD contributions from simulation, including the $t$-channel signal, according to their theory predictions, while the former is modeled using events from the sideband regions described above.
The fit is performed separately in the \rtwojonet and the \rthreejonet categories. The contribution from QCD multijet events to the \rthreejtwot category is only minor and is neglected. To validate the QCD estimation procedure, this fit is also performed in the \rtwojzerot category, a signal-depleted category that provides a number of background events of this source larger by factors of 10 and 28 for the muon and electron channel, respectively. The entire range of the distributions is fitted and the resulting yields of the QCD multijet contribution are then used in the signal regions in which the requirements $\mtw>50\GeV$ or $\ptmiss>30\GeV$ are applied. In this extrapolation, an uncertainty of 50\% is applied to cover all effects from variations in the shape and rate of this background process. Figure~\ref{fig:QCDestimation} shows the fitted \mtw and \ptmiss distributions in the \rtwojonet, \rthreejonet, and \rtwojzerot categories. Good agreement between the results of the fit and the data is found in the low-\mtw and low-\ptmiss regions, where significant contributions from the QCD multijet background are expected. This simple two-template fit is designed to give a reliable estimate of the QCD multijet contribution and is not expected to describe also the tails of the fitted distributions with the same accuracy.

\section{Signal extraction}
\label{sec:sigextract}

\begin{figure*}
    \centering
        \includegraphics[width=0.45\textwidth]{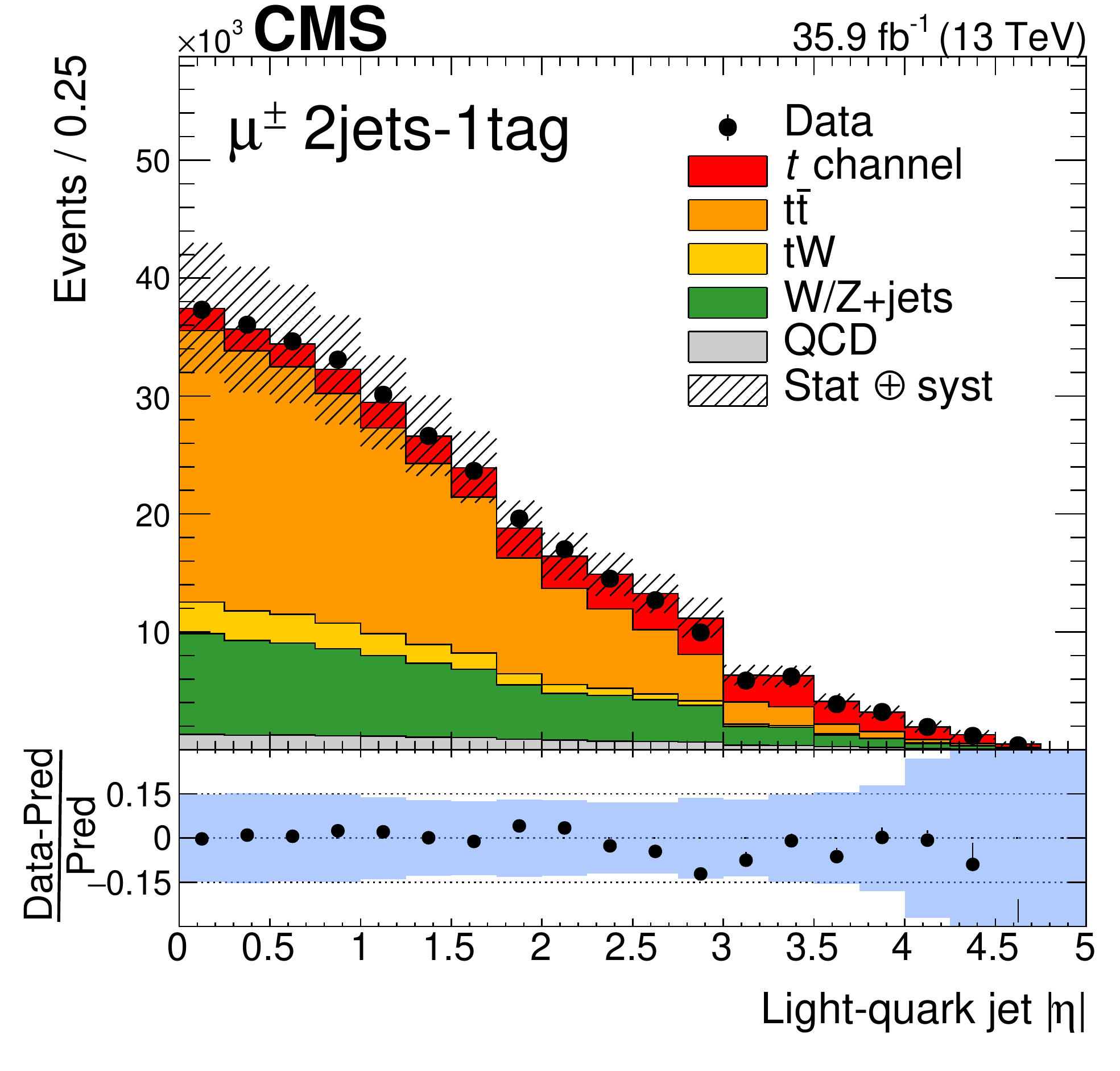}
        \includegraphics[width=0.45\textwidth]{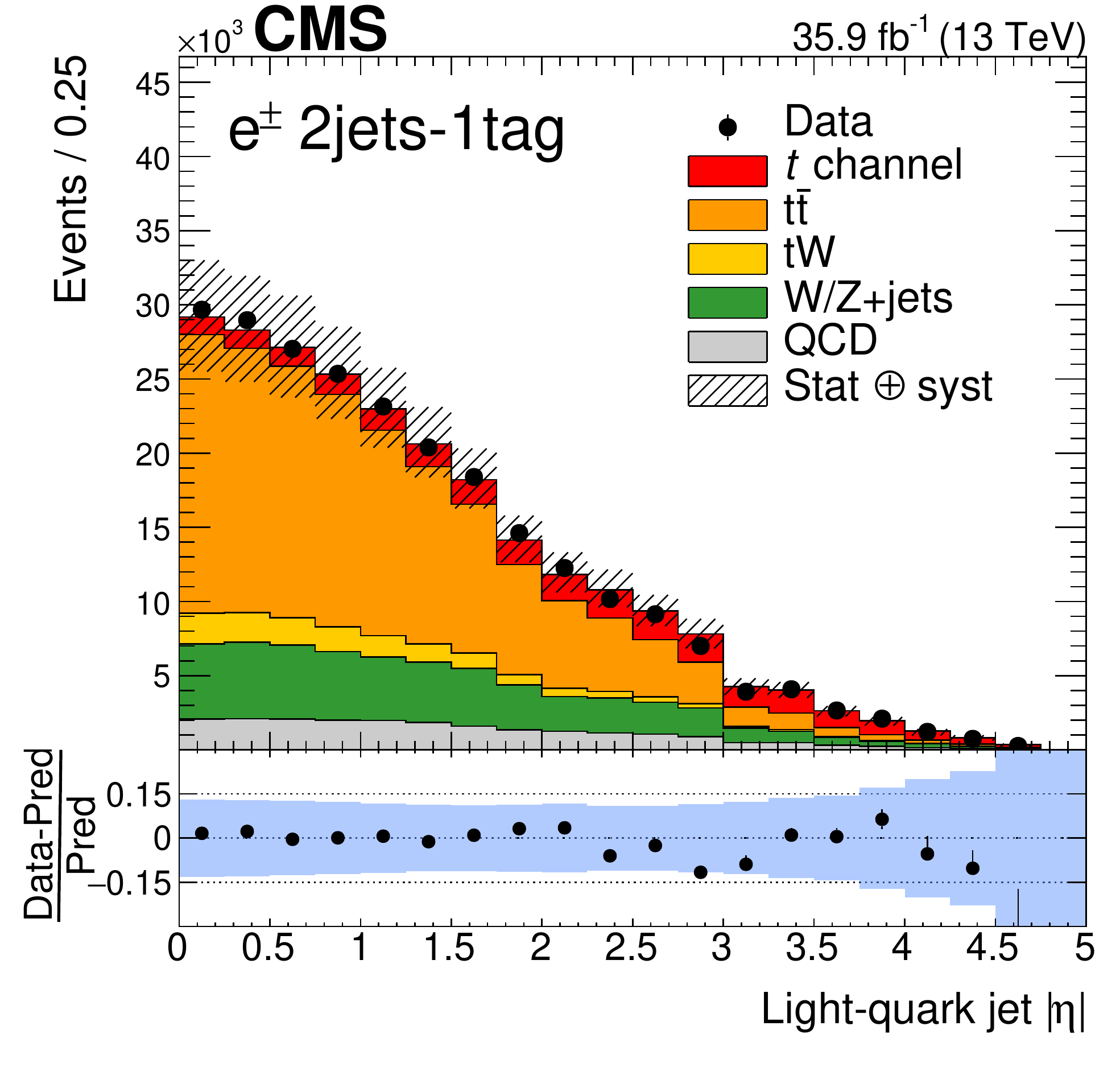} \\
        \includegraphics[width=0.45\textwidth]{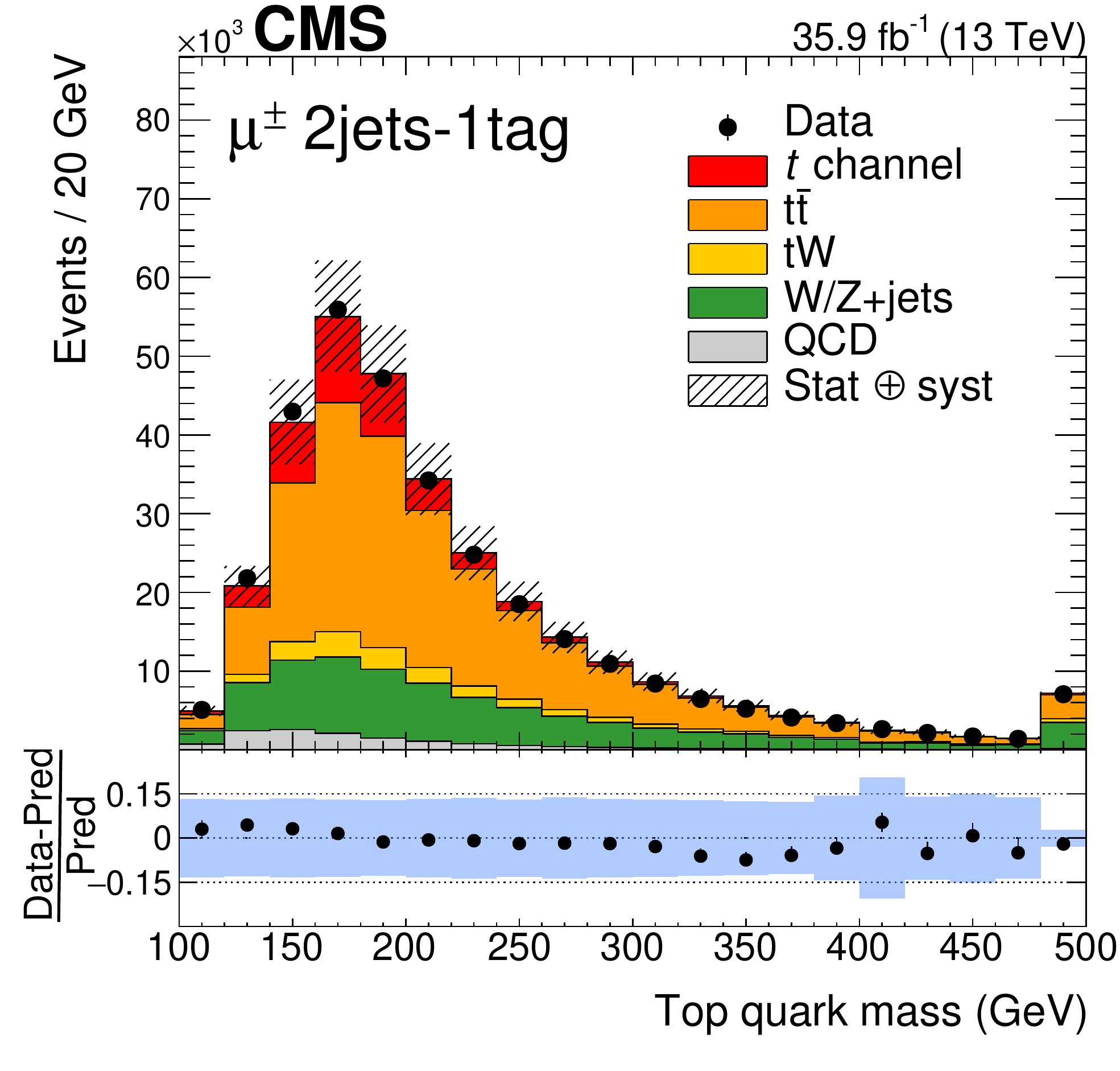}
        \includegraphics[width=0.45\textwidth]{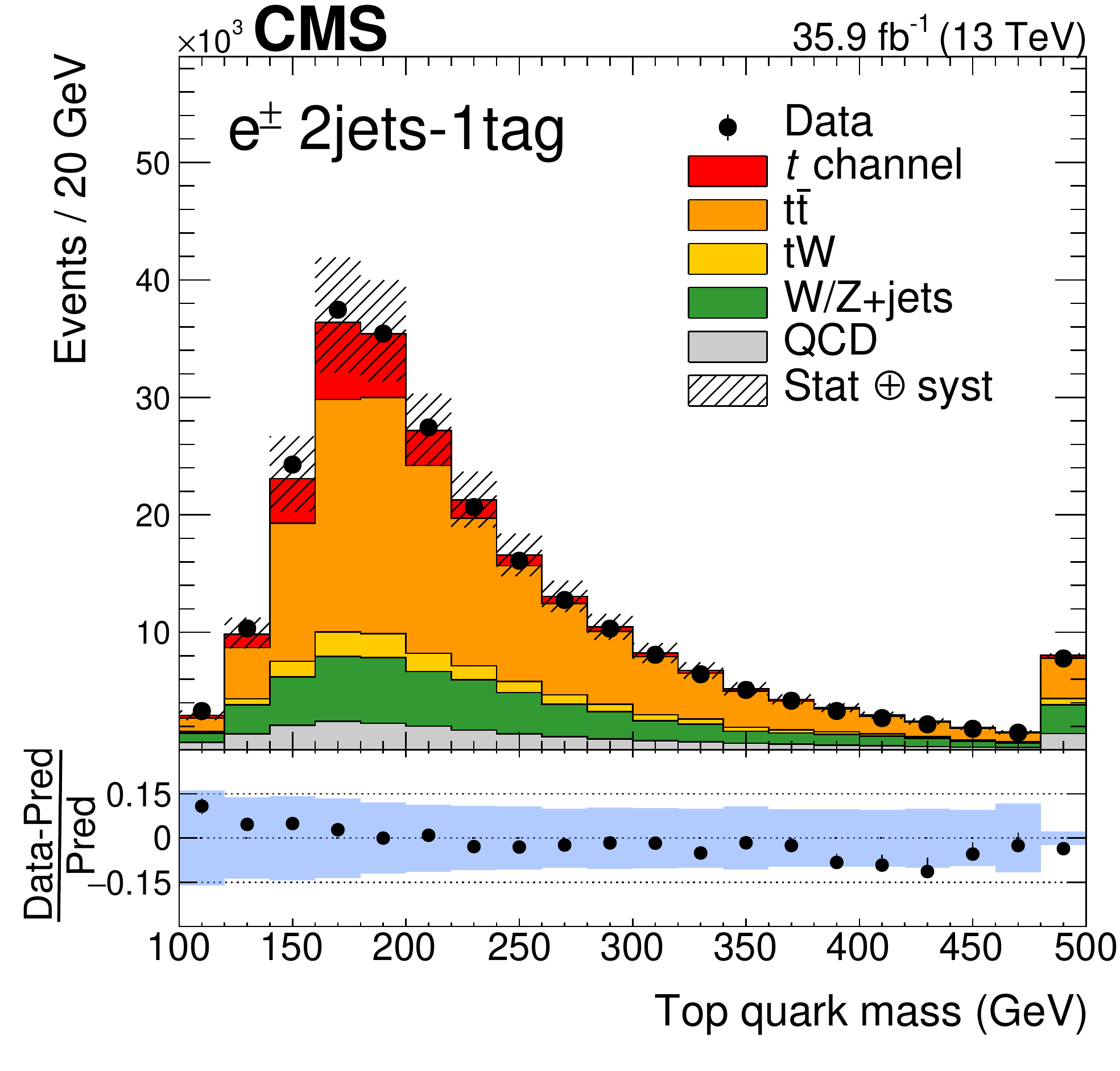} \\
        \includegraphics[width=0.45\textwidth]{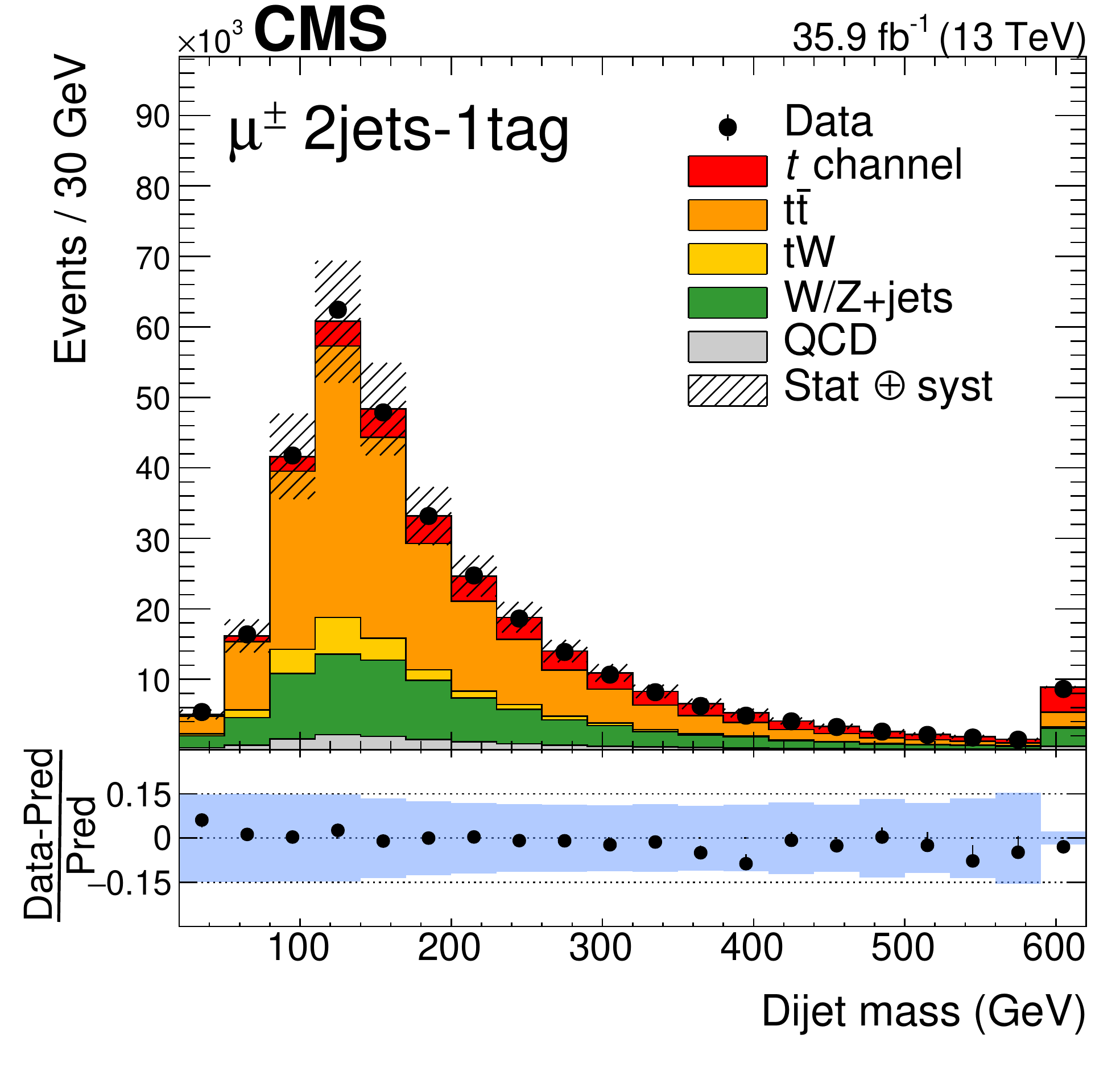}
        \includegraphics[width=0.45\textwidth]{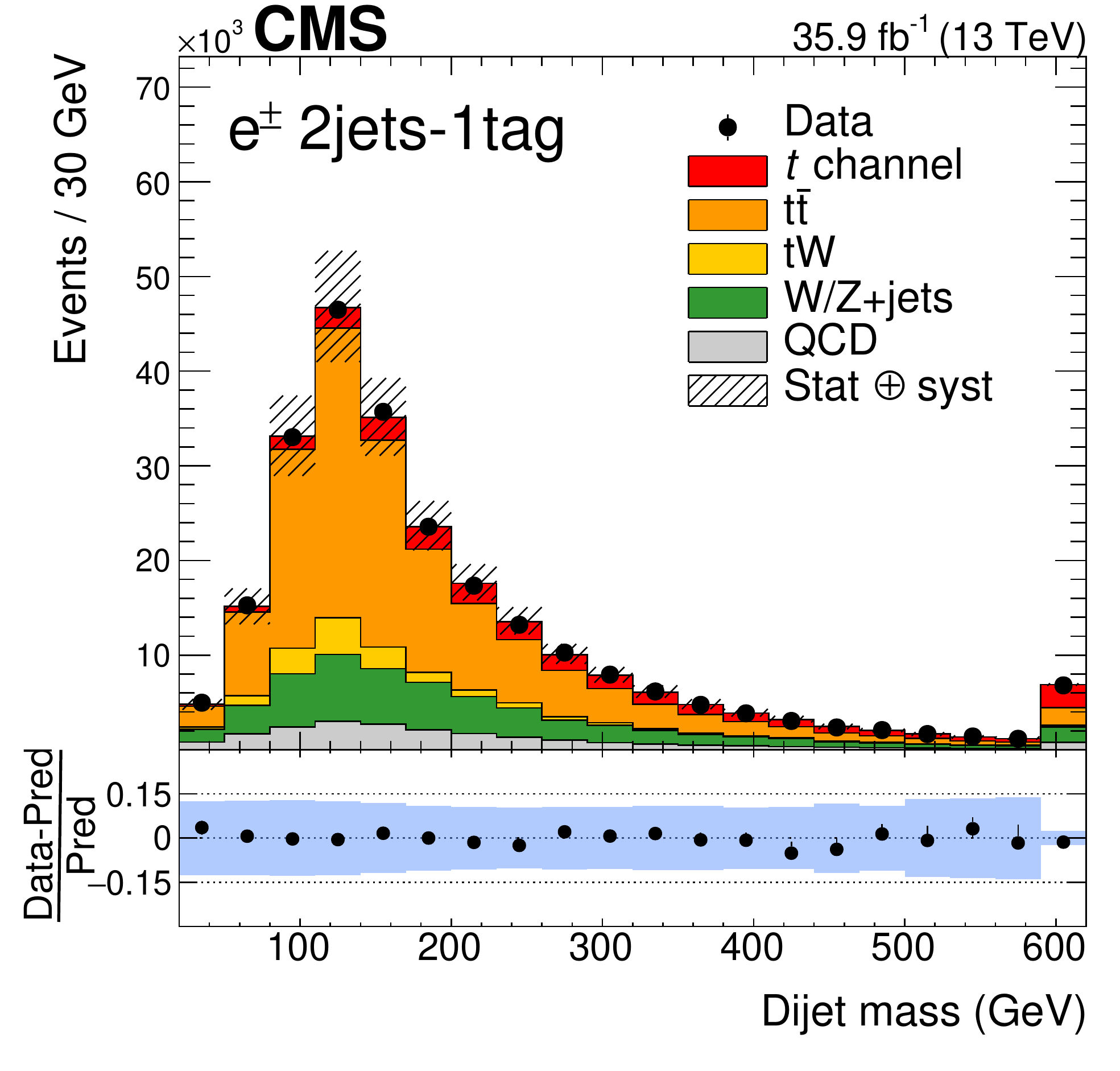}
        \caption{\label{fig:BDTInputVariables1}
        The three most discriminating input variables for the training of the BDTs in the muon channel (left) and in the electron channel (right): the absolute value of the pseudorapidity of the light-quark jet, the mass of the reconstructed top quark, the mass of the light-quark jet and the \cPqb-tagged jet associated to the top quark decay. The variables are ordered by their importance. The simulation is normalized to the total number of events in data. The shaded ares correspond to the quadratic sum of statistical and systematic uncertainties in the simulation before performing the fit. Also shown is the relative difference between the distributions in data and simulation in the lower panels.}
\end{figure*}

\begin{figure*}
    \centering
        \includegraphics[width=0.45\textwidth]{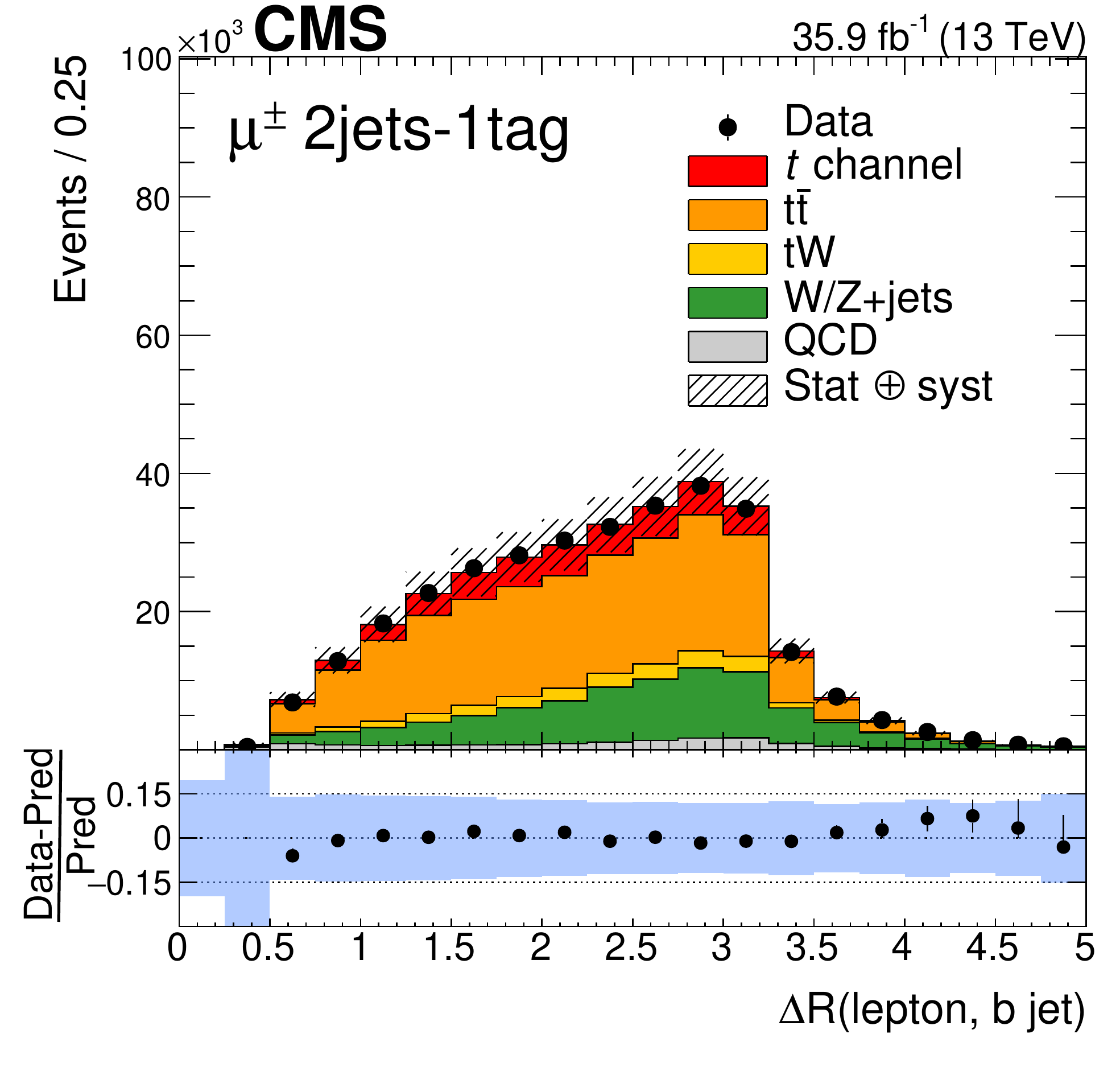}
        \includegraphics[width=0.45\textwidth]{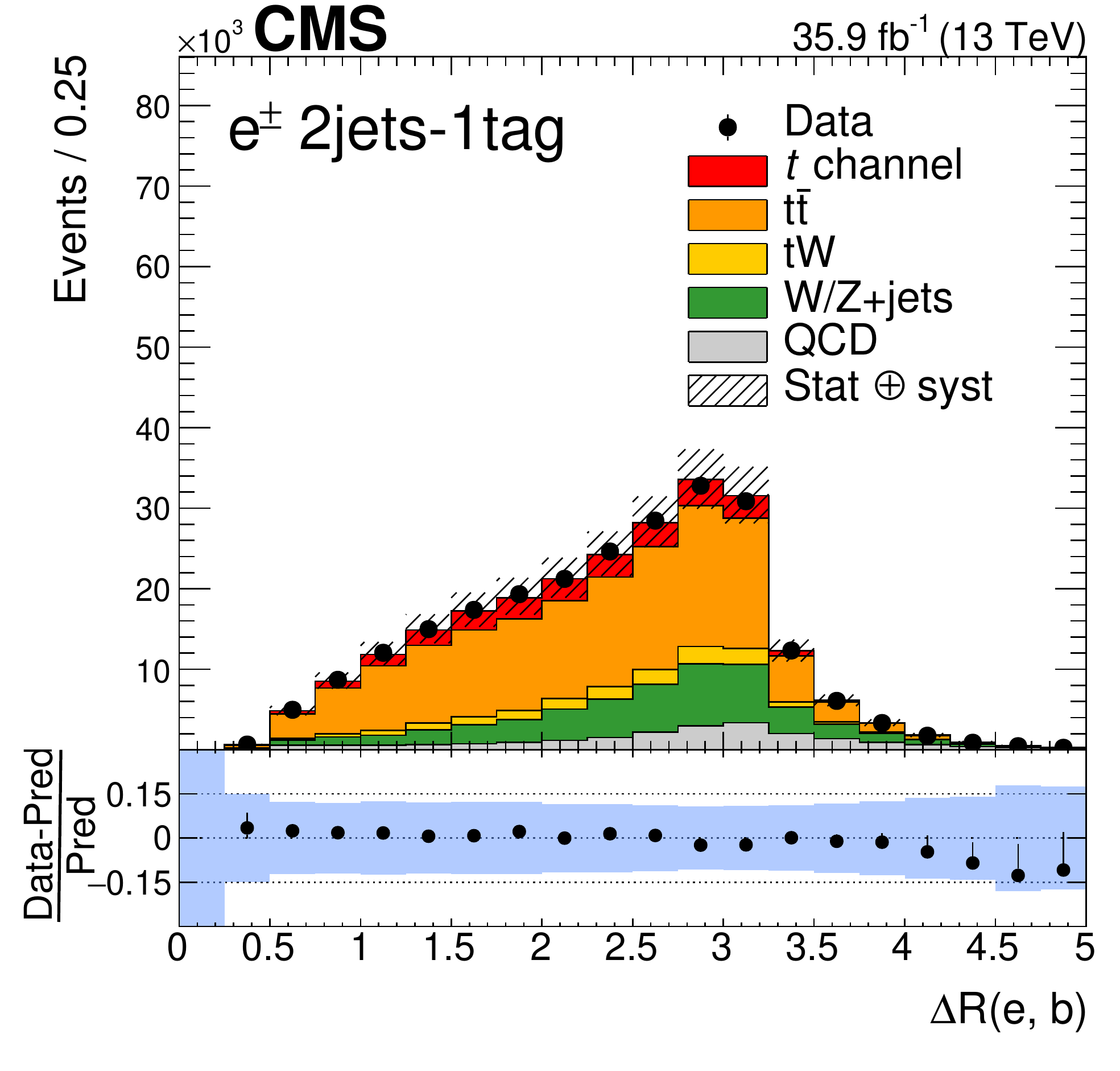} \\
        \includegraphics[width=0.45\textwidth]{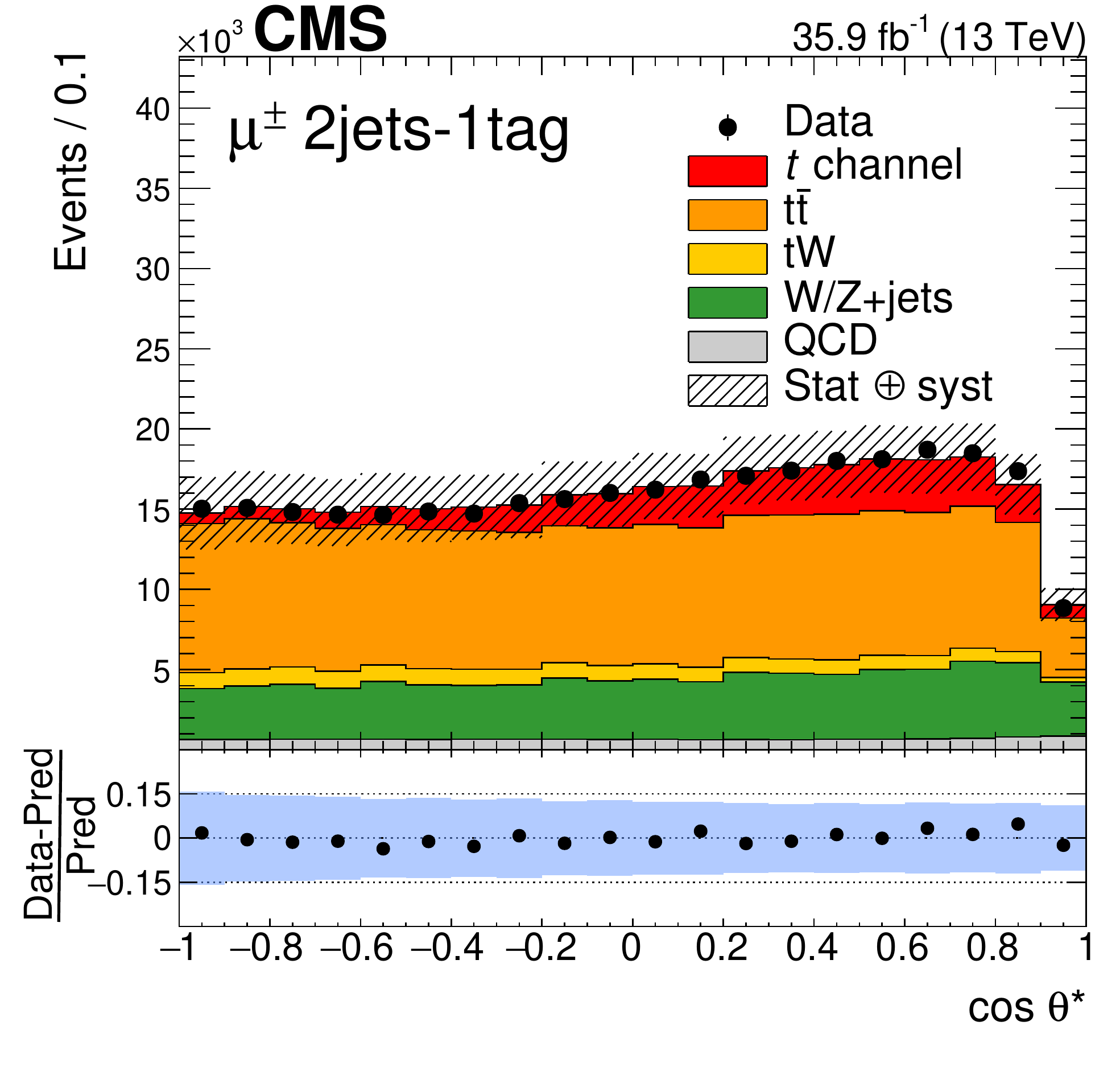}
        \includegraphics[width=0.45\textwidth]{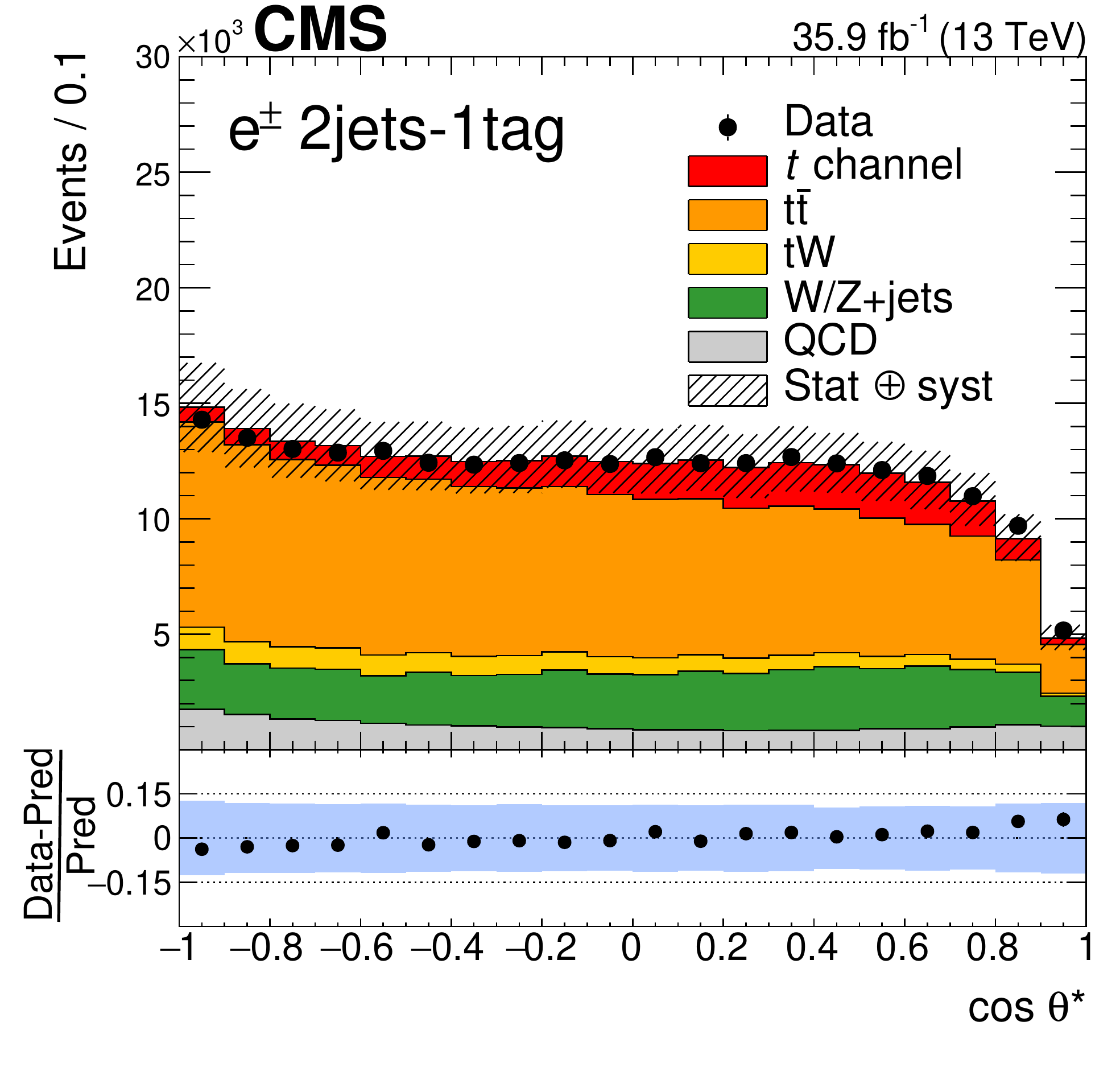}
        \caption{\label{fig:BDTInputVariables2}
        The fourth and fifth most discriminating input variables for the training of the BDTs in the muon channel (left) and in the electron channel (right): the $\Delta{R}$ between the momentum vectors of the lepton and the \cPqb-tagged jet associated with the top quark decay, the cosine of the angle between the lepton and the light-quark jet in the rest frame of the top quark. The simulation is normalized to the total number of events in data. The shaded ares correspond to the quadratic sum of statistical and systematic uncertainties in the simulation before performing the fit. Also shown is the relative difference between the distributions in data and simulation in the lower panels. }
\end{figure*}

BDT algorithms, implemented in the \textsc{tmva} package~\cite{Hocker:2007ht} are employed to combine multiple variables into single discriminators, and thus enhance the separation between signal and background processes. Kinematic variables that are suitable to distinguish the single top quark $t$-channel signal process from the main background contributions are used for the BDT training. Each of these variables is required to be modeled reasonably in the simulation. The list of variables used for discrimination can be found in Table~\ref{tab:ExplanationOFBDTVariables}. The five most important variables are the light-quark jet $\abs{\eta}$, the reconstructed top quark mass, which has high discrimination power against background processes where no top quarks are produced, the invariant mass of the dijet system consisting of the light-quark jet and the \cPqb-tagged jet from the top quark decay, the distance in the $\eta$--$\phi$ plane ($\Delta{R}$) between the charged lepton and the \cPqb\ jet, and the cosine of the angle between the charged lepton and the light-quark jet in the rest frame of the top quark (\costhetastar).
Figures~\ref{fig:BDTInputVariables1} and~\ref{fig:BDTInputVariables2} show the distributions of these five input variables from data compared to the simulation.

The BDTs are trained in the \rtwojonet category, separately for muons and electrons. The lepton $\abs{\eta}$ and \mtw distributions are only considered in events with muons, while the \ptmiss variable is only used in the electron sample. The samples of simulated signal and background events, as well as the QCD multijet sample from sideband data, are normalized according to the respective predictions, with each sample split into two parts. One half is used for the training, while the other half serves for validation purposes and the actual measurement. The trained BDTs are then applied to the \rtwojonet, \rthreejonet, and \rthreejtwot categories, separately for the two different flavors and charges of the lepton.

\begin{figure*}[hbpt!]
    \centering
        \includegraphics[width=0.45\textwidth]{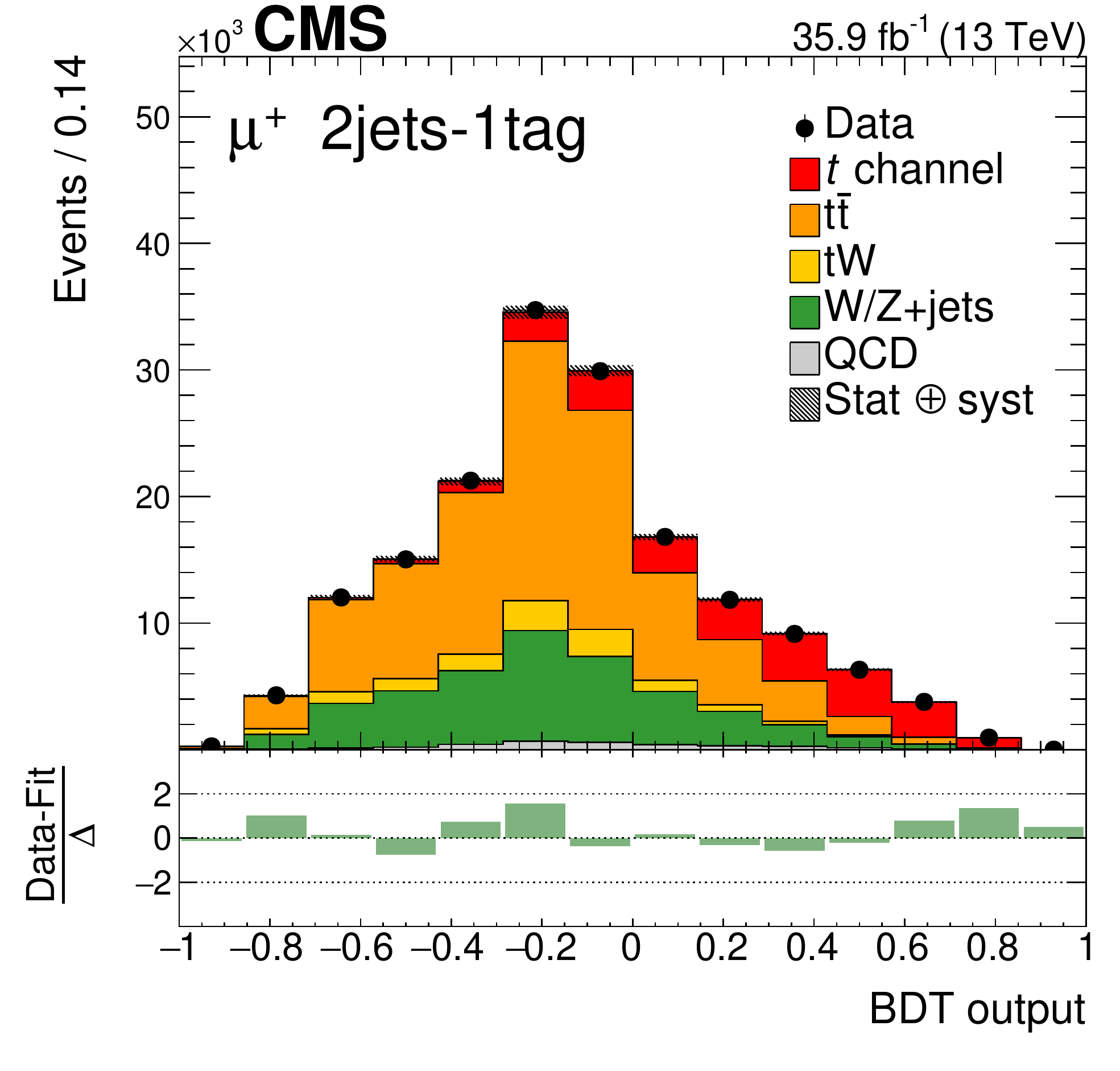}
        \includegraphics[width=0.45\textwidth]{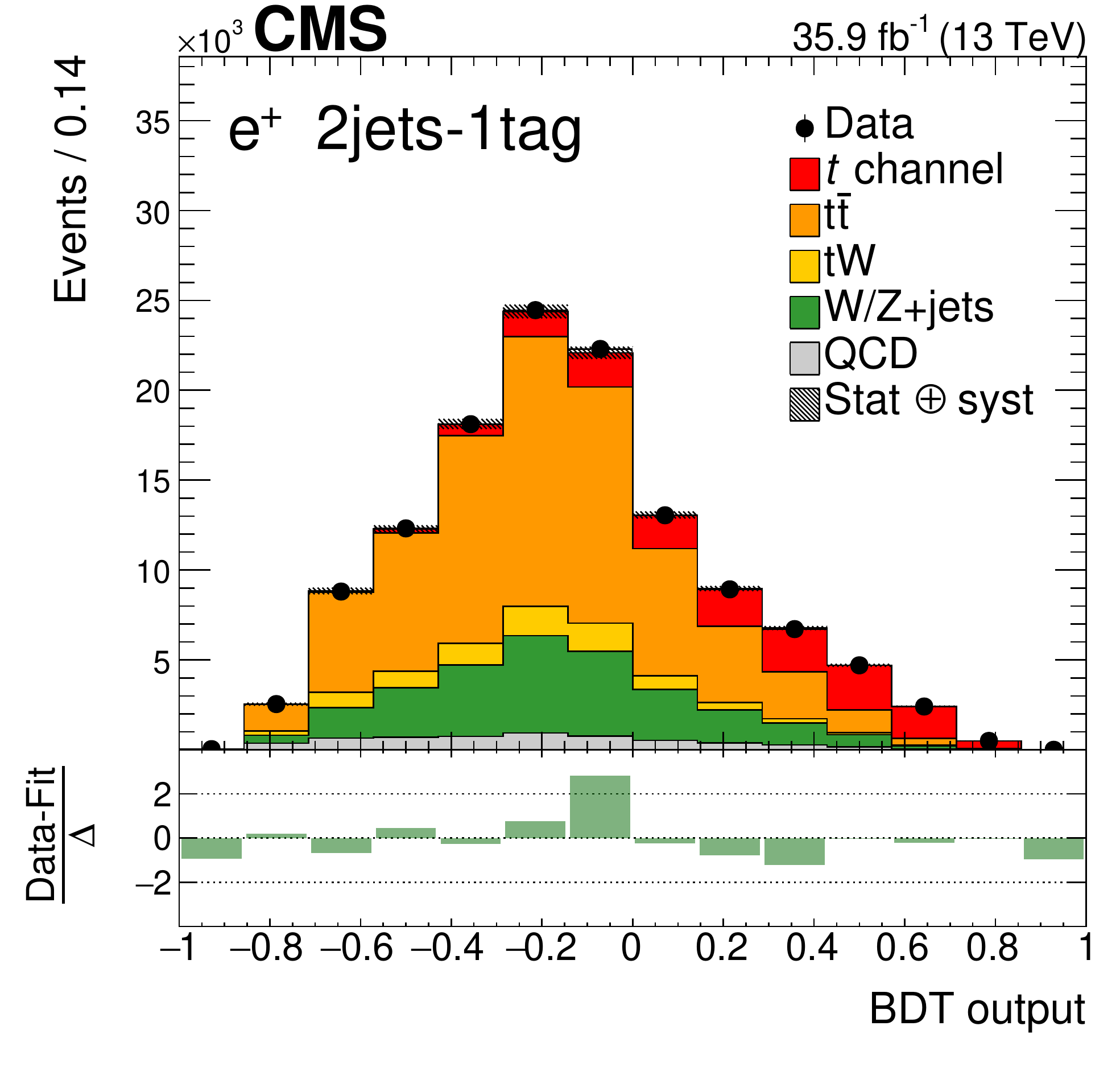}\\
        \includegraphics[width=0.45\textwidth]{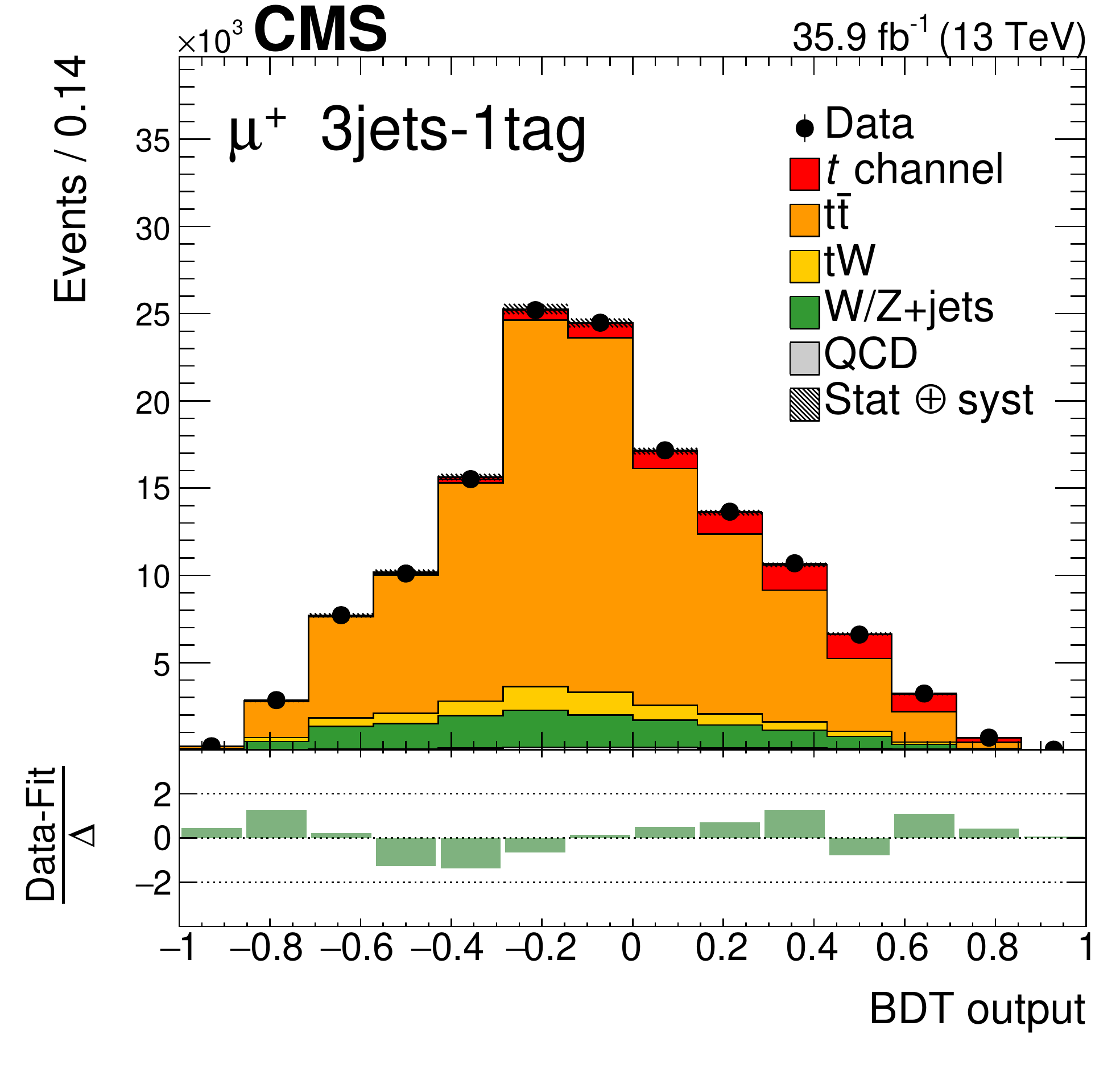}
        \includegraphics[width=0.45\textwidth]{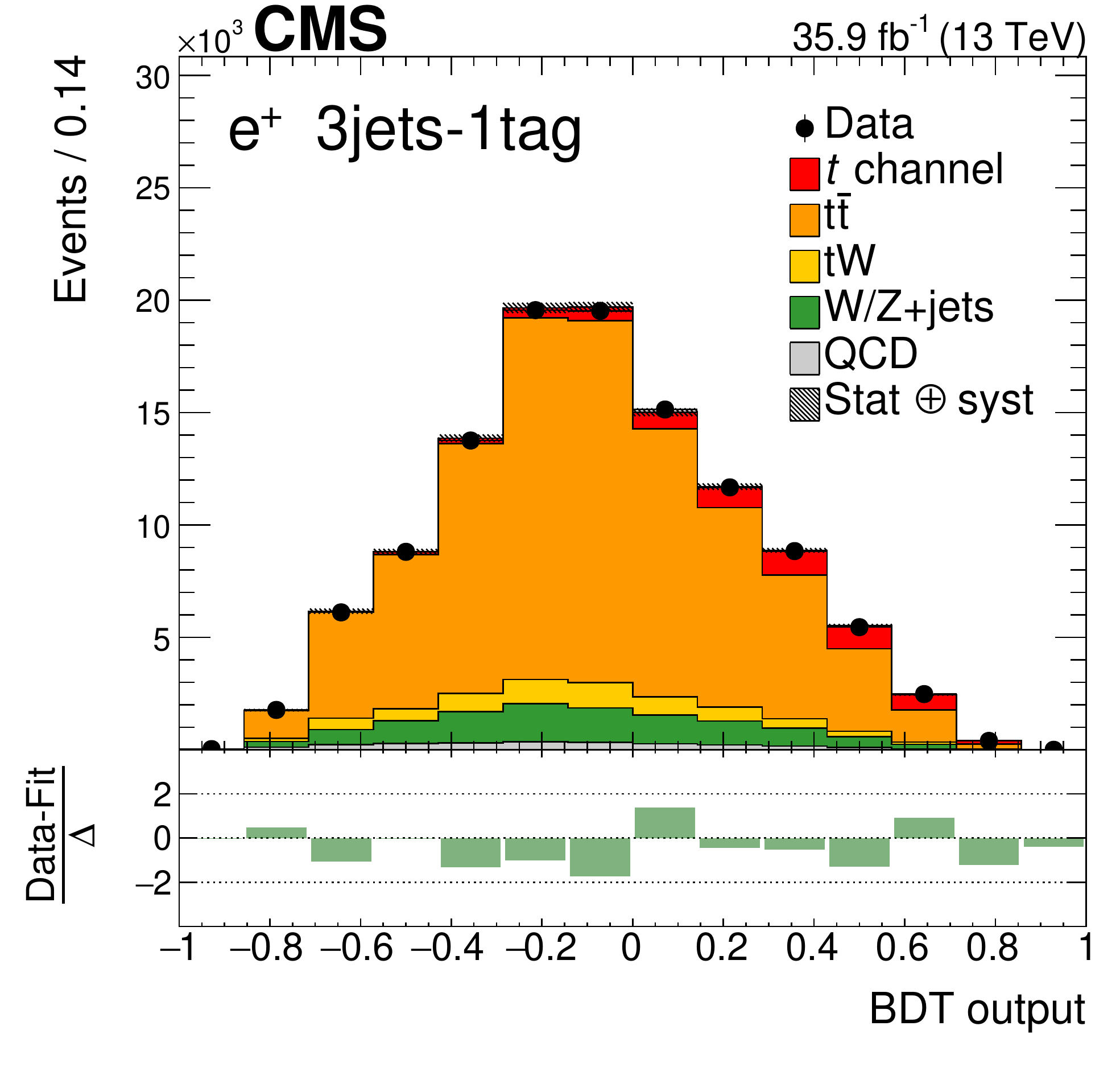} \\
        \includegraphics[width=0.45\textwidth]{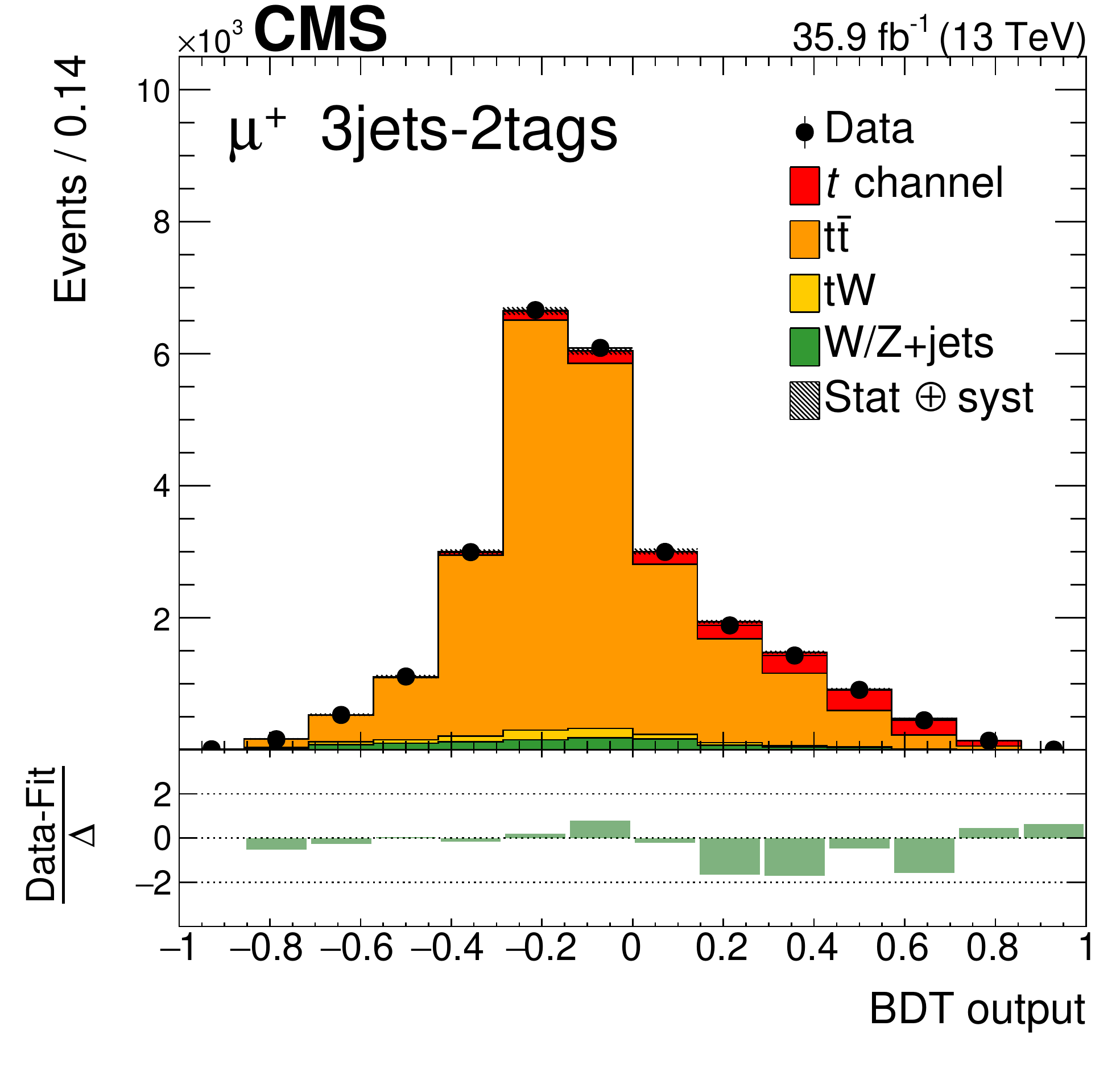}
        \includegraphics[width=0.45\textwidth]{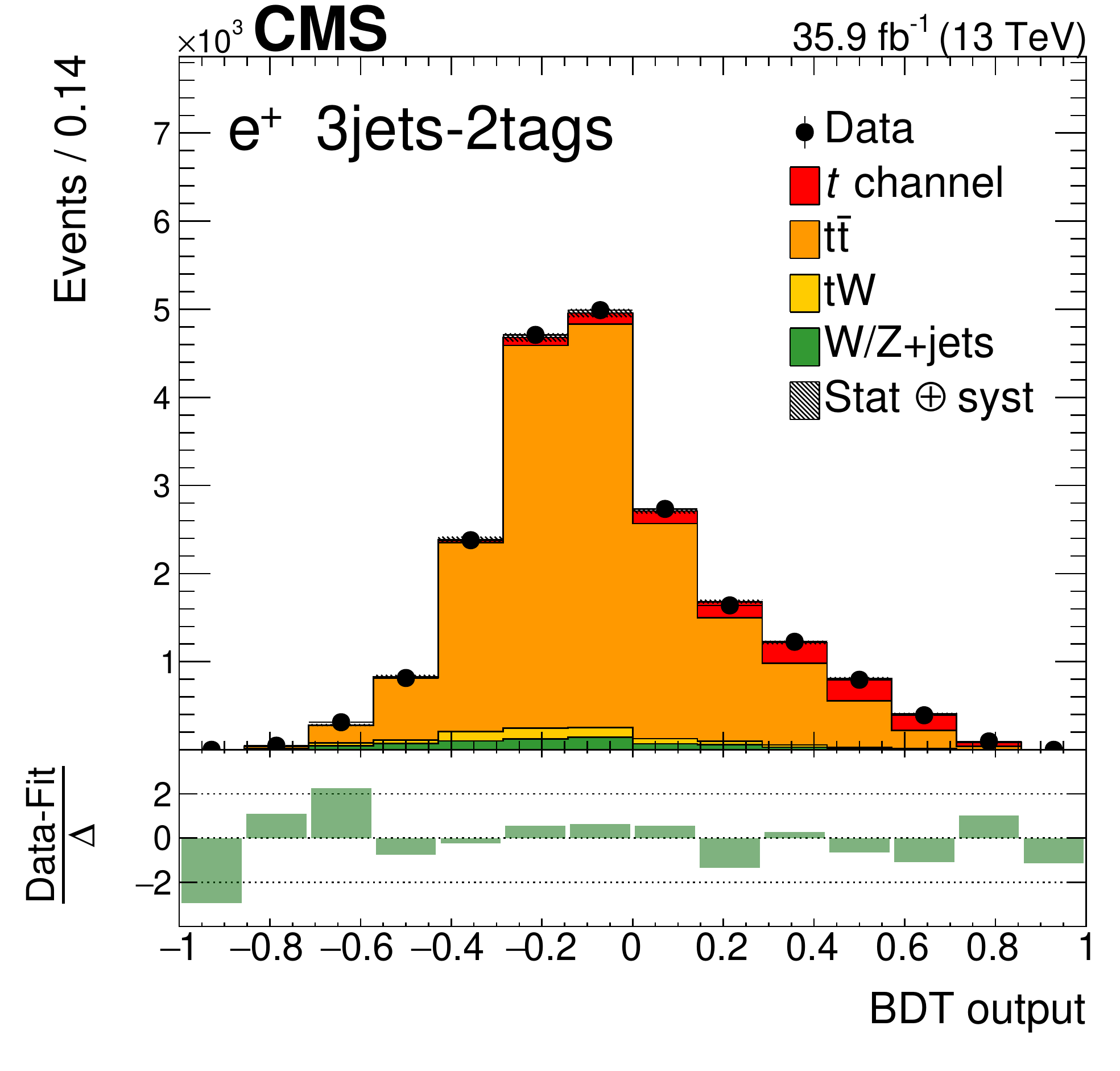}
         \caption{\label{fig:PostFitMVA_pos}
                The BDT output distributions in the \rtwojonet category (upper row), the \rthreejonet category (middle row), and the \rthreejtwot category (lower row) for positively charged muons (left column) and electrons (right column). The different processes are scaled to the corresponding fit results. The shaded areas correspond to the uncertainties after performing the fit. In each figure, the pull is also shown.}
\end{figure*}

\begin{figure*}[hbpt!]
    \centering
        \includegraphics[width=0.45\textwidth]{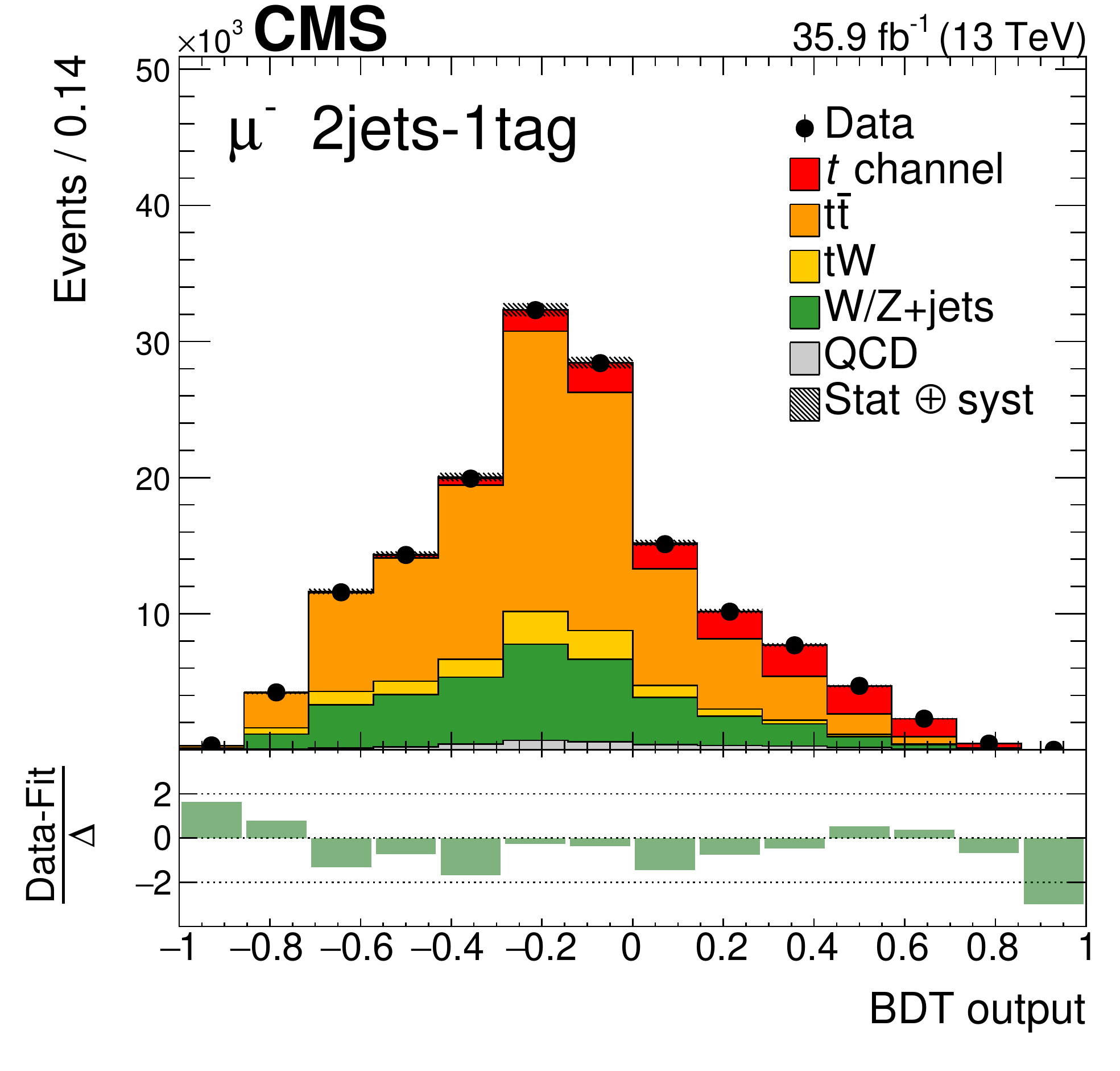}
        \includegraphics[width=0.45\textwidth]{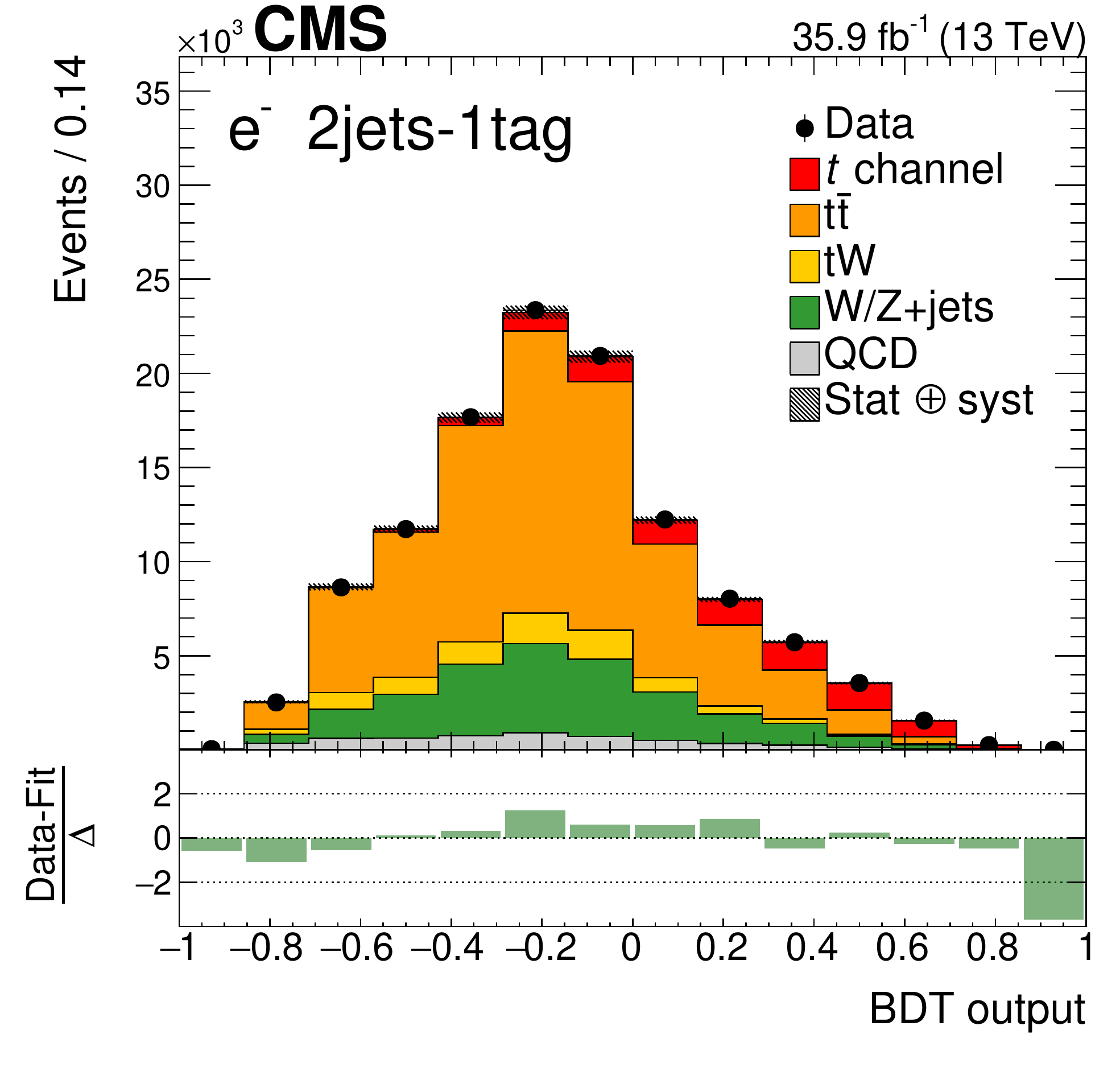}\\
        \includegraphics[width=0.45\textwidth]{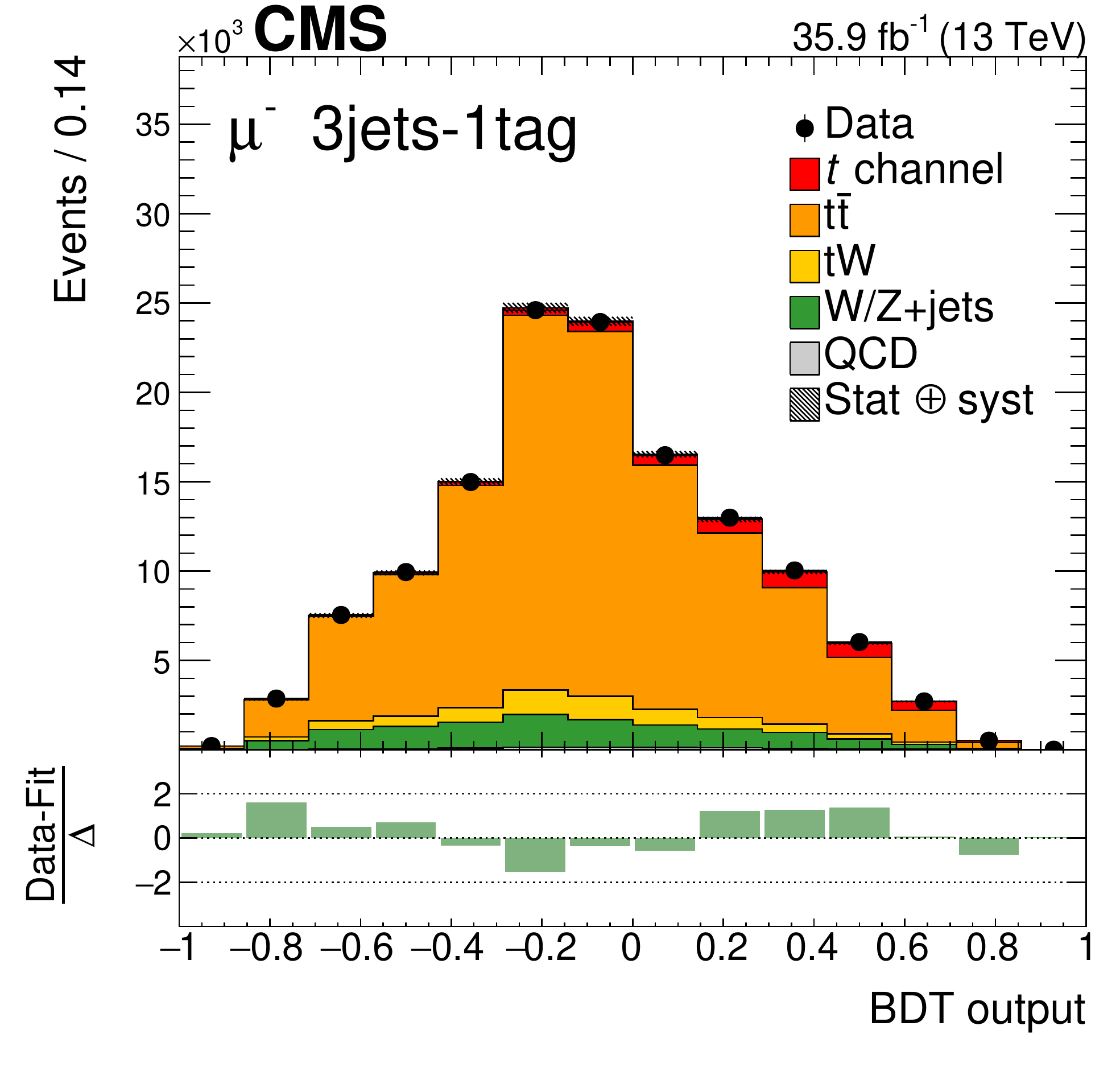}
        \includegraphics[width=0.45\textwidth]{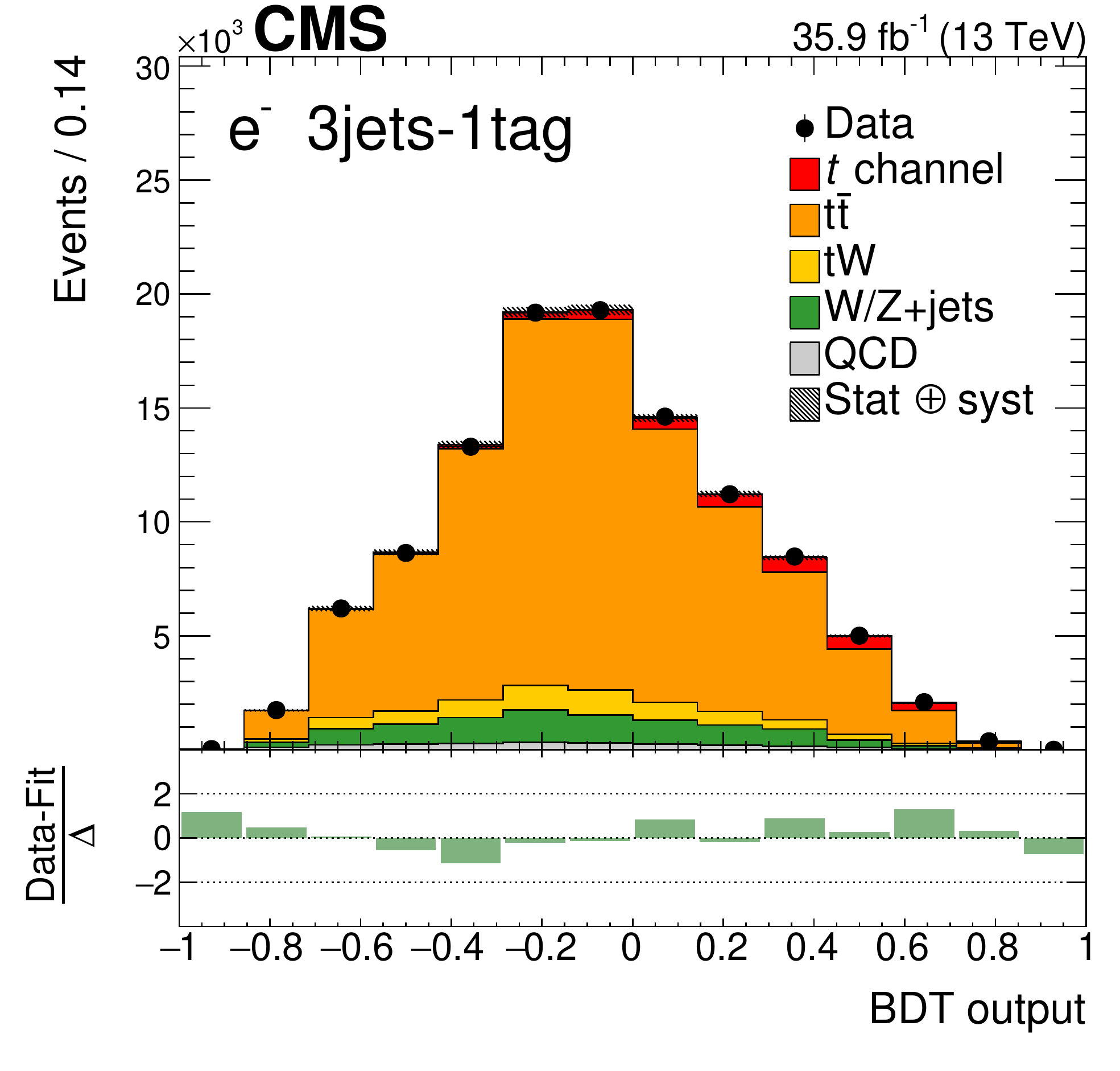} \\
        \includegraphics[width=0.45\textwidth]{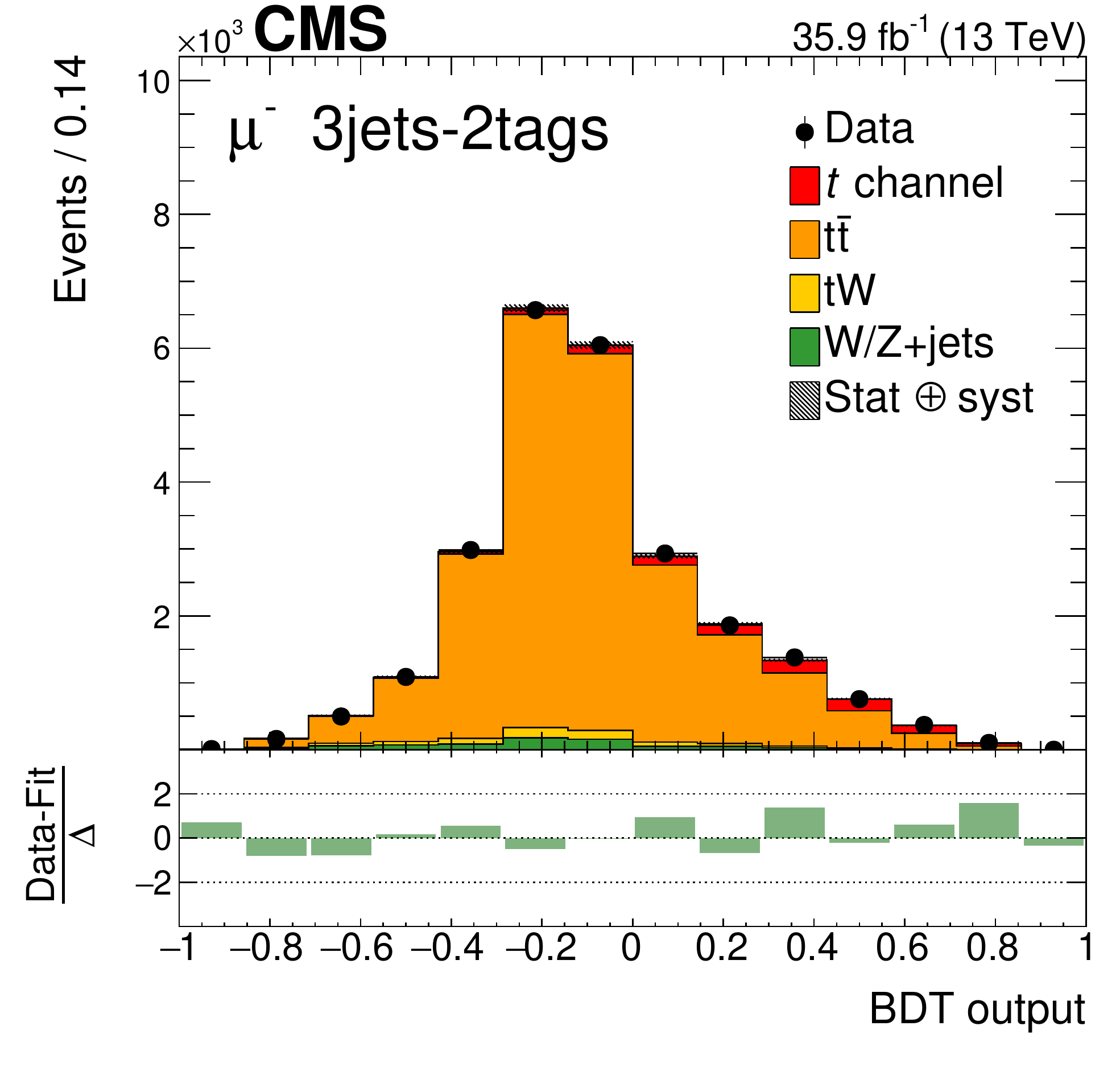}
        \includegraphics[width=0.45\textwidth]{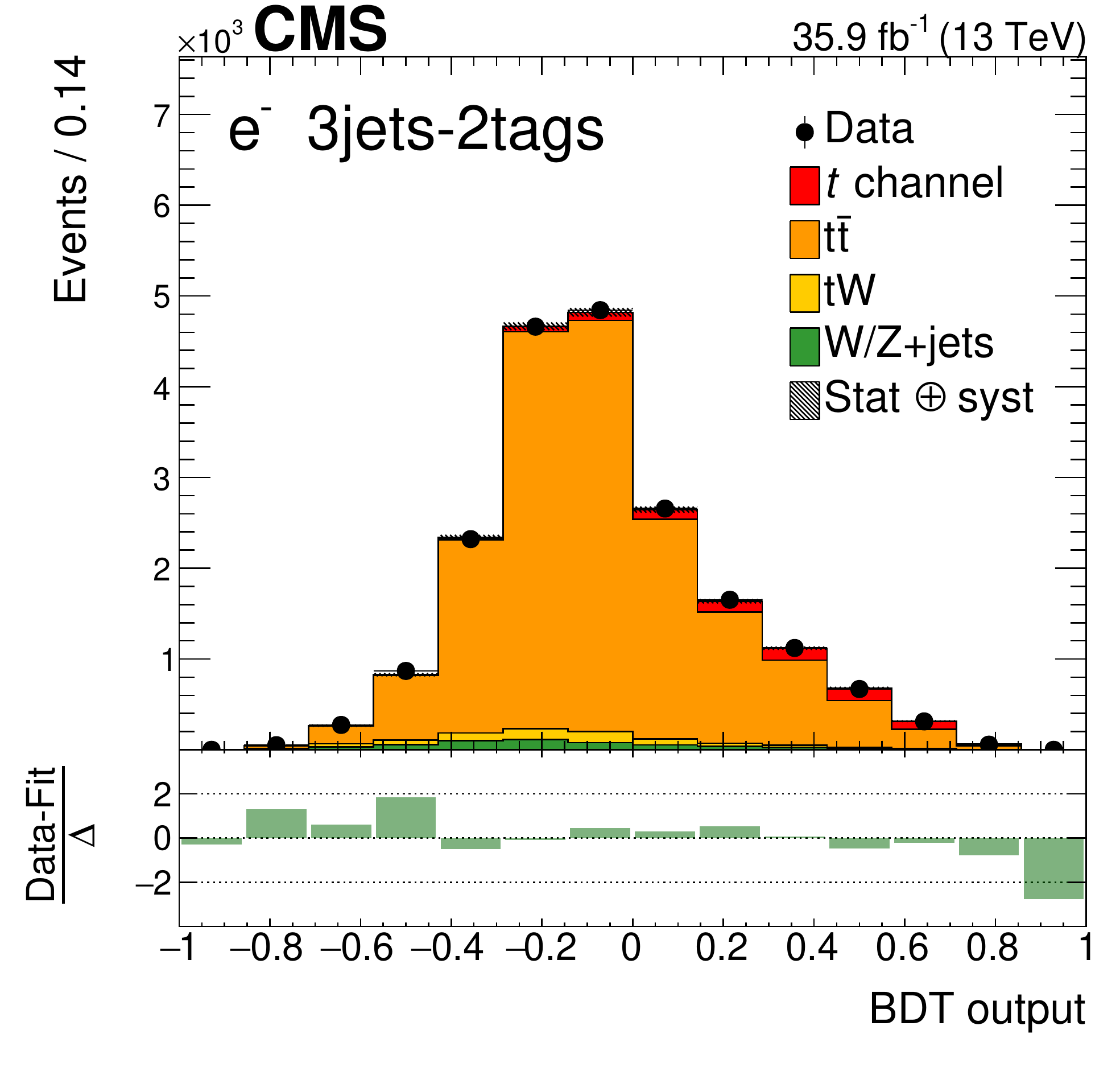}
         \caption{\label{fig:PostFitMVA_neg}
                BDT output distributions in the \rtwojonet category (upper row), the \rthreejonet category (middle row), and the \rthreejtwot category (lower row) for negatively charged muons (left column) and electrons (right column). The different processes are scaled to the corresponding fit results. The shaded areas correspond to the uncertainties after performing the fit. In each figure, the pull is also shown.}
\end{figure*}

A maximum likelihood fit is performed simultaneously on twelve different BDT output distributions (two lepton charges, two lepton flavors, three $n$jets-$m$tags categories). By including the categories with three selected jets in the fit, the \ttbar background, which dominates these categories, is constrained. In this fit, the background rates are determined by introducing nuisance parameters, while the signal rate is a free parameter of the fit. The results of the fit are the cross sections for the production of single top quarks ($\sigmatchtop$) and antiquarks ($\sigmatchantitop$). The ratio of the two cross sections can be calculated from these two results propagating their uncertainties to the ratio using the covariance matrix of the fit. However, a more elegant and straightforward way of properly accounting for the correlations between the uncertainties in the two cross sections is used by repeating the fit with one of the two cross sections replaced by their ratio. This way, potential cancellations of uncertainties are taken into account directly in the fit and do not need to be calculated afterwards. The fitted distributions of the BDT output distributions are shown in Figs.~\ref{fig:PostFitMVA_pos} and~\ref{fig:PostFitMVA_neg}. To verify the quality of the fit, for each distribution, the pull is also shown. The pull is defined as the difference between the distribution in data and the fitted one, divided by the uncertainty $\Delta = \sqrt{\smash[b]{\Delta_{\mathrm{data}}^2 - \Delta_{\mathrm{fit}}^2}}$, where $\Delta_{\mathrm{data}}$ is the Poisson uncertainty in the data and $\Delta_{\mathrm{fit}}$ is the uncertainty of the fit, including the statistical component and all uncertainties that have been included as nuisance parameters. As a cross-check, this fit is also performed separately for each lepton flavor. The obtained values are consistent within their uncertainties compared to the main results in the combined muon and electron channel.

\begin{table*}[h!]
\centering
\topcaption{\label{tab:ExplanationOFBDTVariables}
Input variables for the BDTs. The variables \mtw and lepton $\abs{\eta}$ are only used in the training of events with a muon, while \ptmiss is only considered as input for events with an electron.}
\renewcommand*{\arraystretch}{1.25}
   \begin{tabular}{lp{25em}}
      Variable                           & \multicolumn{1}{l}{Description}                                                                                                        \\ \hline
    Light-quark jet $\abs{\eta}$               & Absolute value of the pseudorapidity of the light-quark jet                                                                            \\
    Top quark mass                     & Invariant mass of the top quark reconstructed from the lepton, the neutrino, and the \cPqb-tagged jet associated to the top quark decay             \\
    Dijet mass                         & Invariant mass of the light-quark jet and the \cPqb-tagged jet associated to the top quark decay                                           \\
    $\Delta{R}$ (lepton, \cPqb\ jet)        & $\Delta{R}$ between the momentum vectors of the lepton and the \cPqb-tagged jet associated with the top quark decay                          \\
    \costhetastar                 & Cosine of the angle between the lepton and the light-quark jet in the rest frame of the top quark                                      \\
    Jet \pt sum           & Scalar sum of the transverse momenta of the light-quark jet and the \cPqb-tagged jet associated to the top quark decay                    \\
    \mtw                               & Transverse mass of the \PW boson                                                                                                         \\
    \ptmiss                            & Missing momentum in the transverse plane of the event                                                                                  \\
    $\Delta{R}$ (light jet, \cPqb\ jet)     & $\Delta{R}$ between the momentum vectors of the light-quark jet and the \cPqb-tagged jet associated to the top quark decay                 \\
    Lepton $\abs{\eta}$                  & Absolute value of the pseudorapidity of the selected lepton                                                                            \\
    \PW boson  $\abs{\eta}$                & Absolute value of the pseudorapidity of the reconstructed \PW boson                                                                      \\
    Light-quark jet mass                     & Invariant mass of the light-quark jet   \\\hline

\end{tabular}
\end{table*}

\section{Systematic uncertainties}
\label{sec:sys}
Several sources of systematic uncertainties are considered in the analysis, either as nuisance parameters in the fit to the BDT distributions (profiled uncertainties) or as nonprofiled uncertainties. While most uncertainty sources, like purely experimental ones and uncertainties in the rates and the modeling of the backgrounds, can be profiled in the fit, this is not possible for those uncertainty sources that are related to the modeling of the signal process. As the results for the signal cross sections are given for the full phase space, the analysis contains an extrapolation from the phase space of the selected data set, in which the actual measurement takes place, to the full phase space, using predictions from the signal simulation about the shapes of the relevant distributions outside the measured area. Hence, the uncertainties in the modeling of the signal process apply not only to the phase space of the selected data set but to the full phase space and should therefore not get constrained from the fit in this reduced phase space. Their impact is determined by repeating the analysis using varied templates according to the systematic uncertainty sources under study in the fit instead of the nominal templates. The larger absolute shift in the parameter of interest, either caused by ``down'' or ``up'' variation of the systematic source under study, is taken as symmetric uncertainty in both directions. In the following, the different uncertainty sources that are considered in the analysis are briefly described. They are grouped in categories of related sources. It is indicated whether the uncertainty sources are implemented via shape morphing (``shape'') or via normalization (``rate'').

\subsection*{Nonprofiled uncertainties}

\begin{itemize}
\item \textit{Signal modeling (shape):}
The following uncertainty sources cover potential mismodeling of the single top quark $t$-channel signal process. They are not considered as nuisance  parameters in the fit but evaluated by repeating the full analysis using samples of simulated signal events that feature variations in the modeling parameters covering the systematic uncertainty sources under study.
\begin{itemize}
\item \textit{Renormalization and factorization scales (shape):}
The uncertainties caused by variations in the renormalization and factorization scales (\murmuf) are considered by applying weights~\cite{Kalogeropoulos:2018cke}, corresponding to simultaneously doubled or halved renormalization and factorization scales with the nominal value set to 172.5\GeV, on the BDT output distributions.
\item \textit{Matching of matrix element and parton shower (shape):}
The parameter $h_\mathrm{damp}=1.581_{-0.585}^{+0.658} \, \mtop$ (with $\mtop = 172.5\GeV$)~\cite{CMS:2016kle}, which controls the matching between the matrix-element level calculation and the parton shower (ME-PS matching) and regulates the high-\pt radiation in the simulation, is varied within its uncertainties.
\item \textit{Parton shower scale (shape):}
The renormalization scales of the initial-state and final-state parton shower (PS) are varied by factors of two and one half with the nominal value set to 172.5\GeV.
\item \textit{Signal PDFs (shape):} The impact due to the choice of PDFs is studied by replacing the nominal signal templates with reweighted distributions derived from the eigenvector variations of NNPDF3.0 NLO~\cite{Ball:2014uwa}.
The full envelope of the eigenvector variations is used.
\end{itemize}
\item \textit{Integrated luminosity (rate):} The relative uncertainty in the integrated luminosity is determined to be $\pm 2.5\%$~\cite{CMS-PAS-LUM-17-001}. This uncertainty is added to the total uncertainties of the measured cross sections.
\end{itemize}

\subsection*{Profiled uncertainties}

\begin{itemize}
\item \textit{Jet energy scale (shape):} All reconstructed jet four-momenta are simultaneously varied in simulation according to the \pt- and $\eta$-dependent uncertainties in the jet energy scale (JES)~\cite{Chatrchyan:2011ds}. In total, 26 different uncorrelated JES uncertainty sources are considered. These variations are also propagated to \ptmiss.
\item \textit{Jet energy resolution (shape):} To account for the difference in the jet energy resolution (JER) between data and simulation, a dedicated smearing is applied to the simulated jets~\cite{Chatrchyan:2011ds}, and the resolutions are varied within their uncertainties.
\item \textit{Unclustered energy (shape):} The contributions of unclustered PF candidates to \ptmiss are varied within their respective energy resolutions~\cite{Khachatryan:2014gga}.
\item \cPqb\ \textit{tagging (shape):} The scale factors that are used to calculate the efficiency corrections of the CSVv2 \cPqb\ tagging algorithm are varied up and down within their uncertainties~\cite{Sirunyan:2017ezt}.
From these up and down varied scale factors, up and down shifted efficiency corrections are calculated and applied to the simulation.
\item \textit{Muon and electron efficiencies (shape):} The efficiencies of the lepton identification and isolation, of the trigger paths, and of the detector response are determined with a ``tag-and-probe'' method~\cite{Khachatryan:2010xn} from Drell--Yan events falling into the \PZ boson mass window. The efficiency correction factors are varied according to the \pt- and $\eta$-dependent uncertainties.
\item \textit{Pileup (shape):} The uncertainty in the average expected number of pileup interactions is propagated as a systematic uncertainty to this measurement by varying the total inelastic cross section by $\pm 4.6\%$~\cite{Sirunyan:2018nqx}.
\item \textit{QCD background normalization (rate):} As described in Section~\ref{sec:bkgmodel}, an uncertainty of $\pm50\%$ is applied to the QCD background estimate to cover all effects from variations in the shape and rate of this process.
\item \textit{Limited size of samples of simulated events (shape):} The limited number of available simulated events is considered by performing the fit using the Barlow--Beeston method~\cite{BARLOW1993219}.
\item \ttbar \textit{background modeling (shape) and normalization (rate):} Multiple systematic effects on the \ttbar{} background prediction are studied: the influence of the parton shower scale and of the matching between the NLO calculation and the parton shower, the impact of variations in the initial- and final-state radiation---depending on the choice of \alpS---and the effect of uncertainties in the modeling of the underlying event. Dedicated \ttbar{} simulation samples are used to study each effect individually, where the corresponding parameters are varied within their uncertainties.
To account for the uncertainty in the prediction of the inclusive \ttbar cross section, a rate uncertainty of $\pm$6\% is applied~\cite{Czakon:2013goa}.
\item \textit{Top quark} \pt \textit{(shape):} In differential measurements of the top quark \pt in \ttbar{} events, the predicted \pt spectrum is found to be harder than the observed spectrum~\cite{Khachatryan:2016mnb,Sirunyan:2017mzl}. To account for this mismodeling, the results derived using the default simulation for \ttbar{} are compared to the results using simulated \ttbar{} events that are reweighted according to the observed difference between data and simulation. This results in one-sided variations of the nominal template.
\item \tW \textit{background normalization (rate):} To account for the uncertainty in the cross section of \tW production and to cover a possible additional systematic uncertainty arising from the procedure which deals with the overlap with the \ttbar{} process at NLO, a rate uncertainty of $\pm11\%$, corresponding to the most precise measurement~\cite{Sirunyan:2018lcp}, is applied. One additional rate uncertainty is included in the fit to account for the impact from the choice of PDFs and their specific variation ($\pm4\%$). To determine the influence of possible mismodeling of the \tW process, the nominal sample is compared to samples generated with a parton shower scale shifted by $\pm1$ standard deviation.
\item $\PW/\PZ$\textit{+jets background normalization (rate):}
To take the uncertainty in the cross sections of the $\PW$+jets and $\PZ$+jets contributions into account, as well as possible effects due to selecting heavy-flavored jets, individual rate uncertainties of $\pm$10\% are applied. By employing these uncertainties, a full evaluation of uncertainty sources for these processes is achieved, as well as a consistent treatment among the different background contributions.
\item \textit{Renormalization and factorization scales (shape):} For the background contributions from \ttbar{}, \tW, and $\PW/\PZ$+jets production, the uncertainties caused by variations in the renormalization and factorization scales (\murmuf) are considered by reweighting~\cite{Kalogeropoulos:2018cke} the BDT output distributions according to simultaneously doubled or halved renormalization and factorization scales. In the case of the \ttbar and \tW processes, the nominal value of the scales is set to 172.5\GeV, while dynamic scales, defined as $\sqrt{\smash[b]{m_{\PW/\PZ}^2 + (\pt^{\PW/\PZ})^2}}$, are used for $\PW/\PZ$+jets production. This uncertainty is estimated for each process separately.
\item \textit{Background PDFs (shape):} By reweighting distributions derived from the eigenvector variations of NNPDF3.0 NLO~\cite{Ball:2014uwa}, the impact due to the choice of PDFs is studied for the \ttbar and $\PW/\PZ$+jets background processes.
The full envelope of the eigenvector variations is used.
The variations in the background PDFs are treated as uncorrelated to the variations in the signal PDFs, because the dominant contributions to the signal PDFs stem from up and down quarks, whereas the background PDFs are dominated by gluons.
\end{itemize}

The by far largest contribution to the selected data set comes from the \ttbar background. To understand this background as well as possible, the fit is performed simultaneously in different $n$jets-$m$tags categories. As a consequence, nuisance parameters for systematic uncertainties that can cause migrations of events between the different categories, like the \ttbar modeling and the jet reconstruction uncertainties, get constrained by the fit. The impact of the individual systematic uncertainties on the measured cross sections and their ratio are listed in Table~\ref{tab:impact_table}. For nonprofiled uncertainties, the change of the result due to the respective variation is listed. The impact of each profiled uncertainty is defined as the shift in the parameter of interest that is induced by repeating the fit with the corresponding nuisance parameter fixed at either one standard deviation above or below its post-fit value, with all other nuisance parameters treated as usual. Of the two resulting shifts always the larger one is taken as the impact. For Table~\ref{tab:impact_table} several nuisance parameters are grouped together by adding their impacts in quadrature.

The dominant uncertainties for the cross section measurements come from variations of the parton shower scale and from the matching between the matrix element and the parton shower employed in the signal modeling. The various uncertainties affect the two cross section measurements in a correlated way, which leads to a significant reduction of their impact when calculating the ratio. The strength of the cancellation depends on the correlation of the respective uncertainties and their impact on the two cross sections. For instance, the nonprofiled signal modeling uncertainties are highly correlated between the two cross sections and the only remaining uncertainty contribution in the ratio comes from the differences in the size of the impacts on the individual cross sections. The dominant uncertainty contributions in the ratio measurement are the uncertainty due to the choice of the PDF set for the $t$-channel signal model and the uncertainty due to the size of the simulation samples.

\begin{table*}
    \centering
      \topcaption{\label{tab:impact_table}Estimated relative impact of uncertainties in percent of the measured cross sections or cross section ratio.}
   \begin{tabular}{ l c c c  }
            & $\Delta R_{t\mathrm{\text{-}ch}}/R_{t\mathrm{\text{-}ch}}$ & $\Delta\sigma/\sigma (\cPqt)$ & $\Delta\sigma/\sigma (\cPaqt)$ \\
        \hline
        \multicolumn{4}{c}{Nonprofiled uncertainties}\\
        \hline
        \murmuf scale $t$ channel & 1.5 & 6.1 & 5.0 \\
        ME-PS scale matching $t$ channel & 0.5 & 7.1 & 7.8 \\
        PS scale $t$ channel & 0.9 & 10.1 & 9.6 \\
        PDF $t$ channel & 3.0 & 3.1 & 5.8 \\
        Luminosity & \NA & 2.5 & 2.5 \\
        \hline
        \multicolumn{4}{c}{Profiled uncertainties}\\
        \hline
	JES & 0.9 & 1.5 & 1.8 \\
	JER & 0.2 & $< 0.1$ & 0.2 \\
	Unclustered energy & $< 0.1$ & 0.1 & 0.2 \\
	\cPqb\ tagging & 0.1 & 1.1 & 1.2 \\
        Muon and electron efficiencies & 0.2 & 0.8 & 0.6 \\
	Pileup & 0.1 & 0.9 & 1.0 \\
	QCD bkg. normalization & $< 0.1$ & 0.1 & 0.1 \\
        MC sample size & 2.5 & 2.2 & 3.2 \\
	\ttbar bkg. model and normalization & 0.2 & 0.6 & 0.6 \\
	Top quark \pt & $< 0.1$ & $< 0.1$ & $< 0.1$ \\
	\tW bkg. normalization & 0.1 & 0.5 & 0.6 \\
        $\PW/\PZ$+jets bkg. normalization & 0.3 & 0.6 & 0.9 \\
        \murmuf scale \ttbar, \tW, $\PW/\PZ$+jets & 0.1 & 0.2 & 0.3 \\
        PDF \ttbar, $\PW/\PZ$+jets& $< 0.1$ & 0.2 & 0.2 \\
	\hline
       \end{tabular}
\end{table*}

\section{Results}
\label{sec:results}
The measured cross sections for the $t$-channel production of single top quarks and antiquarks are
\begin{linenomath}
\begin{align*}
\ifthenelse{\boolean{cms@external}}
{
\sigmatchtop &= \xstopresult \pm \xstopstat \stat \pm \xstopprofiledunc \, \text{(prof)} \pm \xstopextunc \, \text{(sig-mod)} \\
&\qquad \pm \xstoplumiunc \, \text{(lumi)} \unit{pb}\\
}
{
\sigmatchtop &= \xstopresult \pm \xstopstat \stat \pm \xstopprofiledunc \, \text{(prof)} \pm \xstopextunc \, \text{(sig-mod)} \pm \xstoplumiunc \, \text{(lumi)} \unit{pb}\\
}
&= \xstopresult \pm \xstopstat \stat \pm \xstopsysunc \syst \unit{pb}\\
&= \xstopresult \pm \xstoptotalunc \unit{pb},\\
\ifthenelse{\boolean{cms@external}}
{
\sigmatchantitop &= \xsantitopresult \pm \xsantitopstat \stat \pm \xsantitopprofiledunc \, \text{(prof)} \pm \xsantitopextunc \, \text{(sig-mod)} \\
&\qquad \pm \xsantitoplumiunc \, \text{(lumi)} \unit{pb} \\
}
{
\sigmatchantitop &= \xsantitopresult \pm \xsantitopstat \stat \pm \xsantitopprofiledunc \, \text{(prof)} \pm \xsantitopextunc \, \text{(sig-mod)} \pm \xsantitoplumiunc \, \text{(lumi)} \unit{pb} \\
}
&= \xsantitopresult \pm \xsantitopstat \stat \pm \xsantitopsysunc \syst \unit{pb}\\
&= \xsantitopresult \pm \xsantitoptotalunc \unit{pb}.
\end{align*}
\end{linenomath}
Here, the uncertainty sources that are profiled in the fit, are labeled as ``prof'', the uncertainties on the signal modeling are labeled as ``sig-mod'', and the uncertainty due to the integrated luminosity measurement is labeled as ``lumi''. The total systematic uncertainty is obtained by adding the three uncertainty contributions in quadrature. Adding the $\sigmatchtop$ and $\sigmatchantitop$ results, the total cross section is found to be
\begin{linenomath}
\begin{align*}
\ifthenelse{\boolean{cms@external}}
{
\sigmatchtotal &  = \xstotalresult \pm \xstotalstat \stat \pm \xstotalprofiledunc \, \text{(prof)} \pm \xstotalextunc \, \text{(sig-mod)} \\
& \pm \xstotallumunc \, \text{(lumi)}  \unit{pb} \\
}
{
\sigmatchtotal &  = \xstotalresult \pm \xstotalstat \stat \pm \xstotalprofiledunc \, \text{(prof)} \pm \xstotalextunc \, \text{(sig-mod)} \pm \xstotallumunc \, \text{(lumi)}  \unit{pb} \\
}
& = \xstotalresult \pm \xstotalstat \stat \pm \xstotalsysunc \syst \unit{pb} \\
& = \xstotalresult \pm \xstotaltotalunc  \unit{pb},
\end{align*}
\end{linenomath}
where the statistical uncertainties are treated as uncorrelated and the systematic uncertainties as correlated between the $\sigmatchtop$ and $\sigmatchantitop$ measurements. The total cross section is used to calculate the absolute value of the CKM matrix element \vtb. Neglecting $\abs{V_{\cPqt\cPqd}}$ and $\abs{V_{\cPqt\cPqs}}$ as they are significantly smaller than $\abs{\vtb}$, and assuming that the top quark exclusively decays to a \cPqb{} quark and a \PW{} boson, leads to
\begin{linenomath}
\begin{equation*}
\abs{\flv\vtb} = \sqrt{\frac{\sigmatchtotalvtb}{\sigmatchtotalvtbtheo}},
\end{equation*}
\end{linenomath}
with the predicted SM value $\sigmatchtotalvtbtheo=217.0 \substack{+6.6\\-4.6}\,\text{(scale)} \pm 6.2\,\text{(PDF+}\alpS\text{)} \unit{pb}$~\cite{Aliev:2010zk,Kant:2014oha,Botje:2011sn} assuming $\abs{\vtb}=1$. The anomalous form factor \flv takes the possible presence of an anomalous \PW\cPqt\cPqb coupling into account, with $\flv=1$ for the case in which the Wtb interaction is a left-handed weak SM coupling and $\flv\neq1$ for physics beyond the SM~\cite{AguilarSaavedra:2008zc}. The measured cross section translates to
\begin{linenomath}
\begin{equation*}
\abs{\flv\vtb} = \vtbresult \pm \vtbexpunc \,\text{(exp)} \pm 0.02 \,\text{(theo)}.
\end{equation*}
\end{linenomath}
The first uncertainty considers all uncertainties of the cross section measurement, while the second uncertainty is derived from the uncertainty of the theoretical SM prediction. Assuming the unitarity of the CKM matrix, a lower limit of 0.82 is determined in the Feldman--Cousins unified approach~\cite{Feldman:1997qc} for $\abs{\vtb}$ at 95\% confidence level.

\begin{figure*}[tb]
\centering
\includegraphics[width=0.7\textwidth]{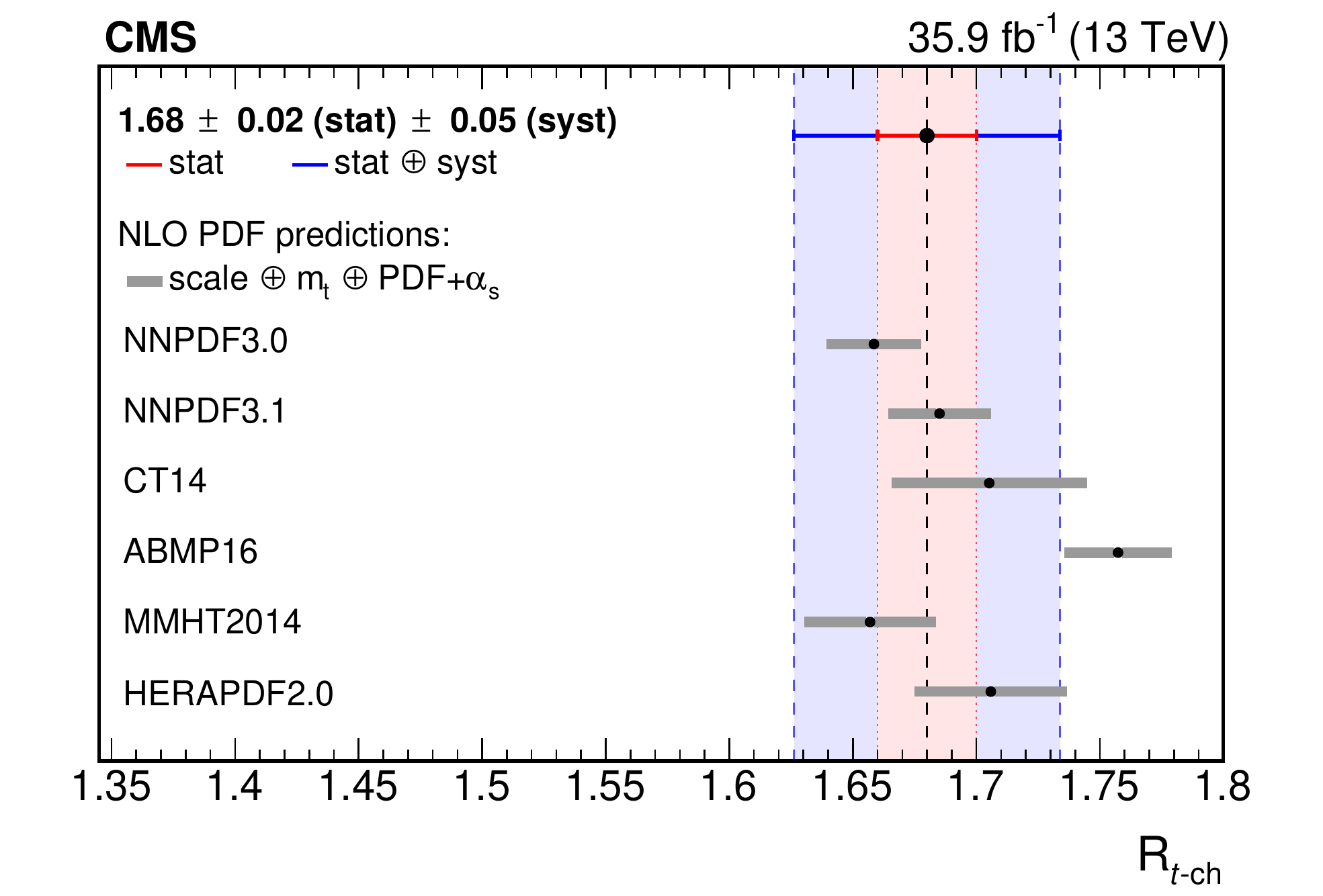}
\caption{\label{fig:Ratio} Comparison of the measured \Rt (central dashed line) with the NLO predictions from different PDF sets, provided by LHAPDF 6.2.1~\cite{Buckley:2014ana}: NNPDF3.0~\cite{Ball:2014uwa}, NNPDF3.1~\cite{Ball:2017nwa}, CT14~\cite{Dulat:2015mca}, ABMP16~\cite{Alekhin:2017kpj,Alekhin:2018}, MMHT2014~\cite{Harland-Lang:2014zoa}, HERAPDF2.0~\cite{Aaron:2009aa}. The \textsc{hathor} 5FS calculation is used with the nominal values for the top quark pole mass and \alpS set to the best values of each PDF set. The uncertainty bars for the different PDF sets include the uncertainty due to the factorization and renormalization scales, the uncertainty in the top quark pole mass, and the combined internal PDF+\alpS uncertainty. For the measurement, the statistical and total uncertainties are indicated individually by the inner and outer uncertainty bars.}
\end{figure*}

The ratio of the cross sections for the production of single top quarks and antiquarks in the $t$ channel is measured as
\begin{linenomath}
\begin{align*}
\Rt &= \rtopantitopresult \pm \rtopantitopstat \, \text{(stat)} \pm \rtopantiprofiledunc \, \text{(prof)} \, \pm \rtopantiextunc \, \text{(sig-mod)} \\
&= \rtopantitopresult \pm \rtopantitopstat \, \text{(stat)} \pm \rtopantisysunc \, \text{(syst)} \\
&= \rtopantitopresult \pm \rtopantitotalunc.
\end{align*}
\end{linenomath}
The measured ratio is compared to the predictions using different PDF sets as shown in Fig.~\ref{fig:Ratio}. Good agreement between the measurement and most predictions is found.

\section{Summary}
\label{sec:summary}
Events with one muon or electron and multiple jets in the final state are used to measure the cross sections for the $t$-channel production of single top quarks and antiquarks, and their ratio. The measured cross sections are $\xstopresult\pm\xstopstat\stat\pm\xstopsysunc\syst\unit{pb}$ for the production of single top quarks, $\xsantitopresult\pm\xsantitopstat\stat\pm\xsantitopsysunc\syst\unit{pb}$ for the production of single top antiquarks, and $\xstotalresult\pm\xstotalstat\stat\pm\xstotalsysunc\syst\unit{pb}$ for the total production. The latter result is used to calculate the absolute value of the Cabibbo--Kobayashi--Maskawa matrix element $\abs{\flv\vtb} = \vtbresult \pm \vtbexpunc \,\text{(exp)} \pm 0.02 \,\text{(theo)}$, including an anomalous form factor $\flv$. The measured ratio of the cross sections of the two processes $\Rt = \rtopantitopresult \pm\rtopantitopstat\stat\pm\rtopantisysunc\syst$ is compared to recent predictions using different parton distribution functions (PDFs) to describe the inner structure of the proton. Good agreement with most PDF sets is found within the uncertainties of the measurement.

The statistical uncertainty plays only a minor role for the achieved precision of the measurements, which are limited by the systematic uncertainties in the modeling of the signal process. Deeper understanding of these effects and improved procedures to estimate the uncertainty are therefore crucial to further decrease the systematic uncertainty. Because of the cancellation of systematic effects when measuring the ratio of cross sections \Rt, its precision, reported in this letter, is significantly improved with respect to the results of previous measurements. The value of \Rt can be used to test the predictions from different PDF sets for their compatibility with the data.

\begin{acknowledgments}
We congratulate our colleagues in the CERN accelerator departments for the excellent performance of the LHC and thank the technical and administrative staffs at CERN and at other CMS institutes for their contributions to the success of the CMS effort. In addition, we gratefully acknowledge the computing centers and personnel of the Worldwide LHC Computing Grid for delivering so effectively the computing infrastructure essential to our analyses. Finally, we acknowledge the enduring support for the construction and operation of the LHC and the CMS detector provided by the following funding agencies: BMBWF and FWF (Austria); FNRS and FWO (Belgium); CNPq, CAPES, FAPERJ, FAPERGS, and FAPESP (Brazil); MES (Bulgaria); CERN; CAS, MoST, and NSFC (China); COLCIENCIAS (Colombia); MSES and CSF (Croatia); RPF (Cyprus); SENESCYT (Ecuador); MoER, ERC IUT, and ERDF (Estonia); Academy of Finland, MEC, and HIP (Finland); CEA and CNRS/IN2P3 (France); BMBF, DFG, and HGF (Germany); GSRT (Greece); NKFIA (Hungary); DAE and DST (India); IPM (Iran); SFI (Ireland); INFN (Italy); MSIP and NRF (Republic of Korea); MES (Latvia); LAS (Lithuania); MOE and UM (Malaysia); BUAP, CINVESTAV, CONACYT, LNS, SEP, and UASLP-FAI (Mexico); MOS (Montenegro); MBIE (New Zealand); PAEC (Pakistan); MSHE and NSC (Poland); FCT (Portugal); JINR (Dubna); MON, RosAtom, RAS, RFBR, and NRC KI (Russia); MESTD (Serbia); SEIDI, CPAN, PCTI, and FEDER (Spain); MOSTR (Sri Lanka); Swiss Funding Agencies (Switzerland); MST (Taipei); ThEPCenter, IPST, STAR, and NSTDA (Thailand); TUBITAK and TAEK (Turkey); NASU and SFFR (Ukraine); STFC (United Kingdom); DOE and NSF (USA).

\hyphenation{Rachada-pisek} Individuals have received support from the Marie-Curie program and the European Research Council and Horizon 2020 Grant, contract No. 675440 (European Union); the Leventis Foundation; the A.P.\ Sloan Foundation; the Alexander von Humboldt Foundation; the Belgian Federal Science Policy Office; the Fonds pour la Formation \`a la Recherche dans l'Industrie et dans l'Agriculture (FRIA-Belgium); the Agentschap voor Innovatie door Wetenschap en Technologie (IWT-Belgium); the F.R.S.-FNRS and FWO (Belgium) under the ``Excellence of Science -- EOS" -- be.h project n.\ 30820817; the Ministry of Education, Youth and Sports (MEYS) of the Czech Republic; the Lend\"ulet (``Momentum") Program and the J\'anos Bolyai Research Scholarship of the Hungarian Academy of Sciences, the New National Excellence Program \'UNKP, the NKFIA research grants 123842, 123959, 124845, 124850, and 125105 (Hungary); the Council of Science and Industrial Research, India; the HOMING PLUS program of the Foundation for Polish Science, cofinanced from European Union, Regional Development Fund, the Mobility Plus program of the Ministry of Science and Higher Education, the National Science Center (Poland), contracts Harmonia 2014/14/M/ST2/00428, Opus 2014/13/B/ST2/02543, 2014/15/B/ST2/03998, and 2015/19/B/ST2/02861, Sonata-bis 2012/07/E/ST2/01406; the National Priorities Research Program by Qatar National Research Fund; the Programa Estatal de Fomento de la Investigaci{\'o}n Cient{\'i}fica y T{\'e}cnica de Excelencia Mar\'{\i}a de Maeztu, grant MDM-2015-0509 and the Programa Severo Ochoa del Principado de Asturias; the Thalis and Aristeia programs cofinanced by EU-ESF and the Greek NSRF; the Rachadapisek Sompot Fund for Postdoctoral Fellowship, Chulalongkorn University and the Chulalongkorn Academic into Its 2nd Century Project Advancement Project (Thailand); the Welch Foundation, contract C-1845; and the Weston Havens Foundation (USA).
\end{acknowledgments}

\bibliography{auto_generated}
\cleardoublepage \appendix\section{The CMS Collaboration \label{app:collab}}\begin{sloppypar}\hyphenpenalty=5000\widowpenalty=500\clubpenalty=5000\vskip\cmsinstskip
\textbf{Yerevan Physics Institute, Yerevan, Armenia}\\*[0pt]
A.M.~Sirunyan, A.~Tumasyan
\vskip\cmsinstskip
\textbf{Institut f\"{u}r Hochenergiephysik, Wien, Austria}\\*[0pt]
W.~Adam, F.~Ambrogi, E.~Asilar, T.~Bergauer, J.~Brandstetter, M.~Dragicevic, J.~Er\"{o}, A.~Escalante~Del~Valle, M.~Flechl, R.~Fr\"{u}hwirth\cmsAuthorMark{1}, V.M.~Ghete, J.~Hrubec, M.~Jeitler\cmsAuthorMark{1}, N.~Krammer, I.~Kr\"{a}tschmer, D.~Liko, T.~Madlener, I.~Mikulec, N.~Rad, H.~Rohringer, J.~Schieck\cmsAuthorMark{1}, R.~Sch\"{o}fbeck, M.~Spanring, D.~Spitzbart, A.~Taurok, W.~Waltenberger, J.~Wittmann, C.-E.~Wulz\cmsAuthorMark{1}, M.~Zarucki
% ...
\vskip\cmsinstskip
% ...
1:  Also at Vienna University of Technology, Vienna, Austria\\
% ...
\end{sloppypar}
\end{document}